\RequirePackage{ifpdf}
\ifpdf 
\documentclass[pdftex]{sigma}
\else
\documentclass{sigma}
\fi

\newcommand{\FUTA}{{\bf FUTA}}
\newcommand{\FUTB}{{\bf FUTB}}
\newcommand{\lsim}{\;\raisebox{-.3em}{$\stackrel{\displaystyle <}{\sim}$}\;}

\numberwithin{equation}{section}

\begin{document}

\allowdisplaybreaks

\renewcommand{\thefootnote}{$\star$}

\renewcommand{\PaperNumber}{049}

\FirstPageHeading

\ShortArticleName{Finite Unif\/ication: Theory and Predictions}

\ArticleName{Finite Unif\/ication: Theory and Predictions\footnote{This
paper is a contribution to the Proceedings of the Eighth
International Conference ``Symmetry in Nonlinear Mathematical
Physics'' (June 21--27, 2009, Kyiv, Ukraine). The full collection
is available at
\href{http://www.emis.de/journals/SIGMA/symmetry2009.html}{http://www.emis.de/journals/SIGMA/symmetry2009.html}}}

\Author{Sven HEINEMEYER~$^{\dag^1}$, Myriam MONDRAG\'ON~$^{\dag^2}$ and George ZOUPANOS~$^{\dag^3\dag^4}$}

\AuthorNameForHeading{S.~Heinemeyer, M. Mondrag\'on and G. Zoupanos}

\Address{$^{\dag^1}$~Instituto de F\'{\i}sica de Cantabria (CSIC-UC), Santander, Spain}
\EmailDD{\href{mailto:Sven.Heinemeyer@cern.ch}{Sven.Heinemeyer@cern.ch}}

\Address{$^{\dag^2}$~Instituto de F\'{\i}sica, Universidad Nacional Aut\'onoma de M\'exico,\\
\hphantom{$^{\dag^2}$}~Apdo.~Postal 20-364, M\'exico 01000, M\'exico}
\EmailDD{\href{mailto:myriam@fisica.unam.mx}{myriam@fisica.unam.mx}}

\Address{$^{\dag^3}$~Theory Group, Physics Department,  CERN, Geneva, Switzerland}
\EmailDD{\href{mailto:George.Zoupanos@cern.ch}{George.Zoupanos@cern.ch}}

\Address{$^{\dag^4}$~Physics Department,  National Technical University, 157 80 Zografou, Athens, Greece}

\ArticleDates{Received January 03, 2010, in f\/inal form May 25, 2010;  Published online June 11, 2010}

\Abstract{All-loop Finite Unif\/ied Theories (FUTs) are very interesting $N=1$
supersymmetric Grand Unif\/ied Theories (GUTs) which not only realise an
old f\/ield theoretic dream but also have a remarkable predictive power due
to the required reduction of couplings. The reduction of the dimensionless
couplings in $N=1$ GUTs is achieved by searching for renormalization group
invariant (RGI) relations among them holding beyond the unif\/ication
scale. Finiteness results from the fact that there exist RGI relations
among dimensionless couplings that guarantee the vanishing of all
beta-functions in certain $N=1$ GUTs even to all orders. Furthermore
developments in the soft supersymmetry breaking sector of $N=1$ GUTs and
FUTs lead to exact RGI relations, i.e.\ reduction of couplings, in this
dimensionful sector of the theory too. Based on the above theoretical
framework phenomenologically consistent FUTS have been constructed. Here
we present FUT models based on the $SU(5)$ and $SU(3)^3$ gauge groups and
their predictions. Of particular interest is the Higgs mass prediction of
one of the models which is expected to be tested at the LHC.}


\Classification{81T60; 81V22}

\renewcommand{\thefootnote}{\arabic{footnote}}
\setcounter{footnote}{0}

\section{Introduction}

A large and sustained ef\/fort has been done in the recent years aiming
to achieve a unif\/ied description of all interactions. Out of this
endeavor two main directions have emerged as the most promising to
attack the problem, namely, the superstring theories and
non-commutative geometry. The two approaches, although at a dif\/ferent
stage of development, have common unif\/ication targets and share similar
hopes for exhibiting improved renormalization properties in the
ultraviolet (UV) as compared to ordinary f\/ield theories.  Moreover the
two frameworks came closer by the observation that a natural
realization of non-commutativity of space appears in the string theory
context of D-branes in the presence of a constant background
antisymmetric f\/ield \cite{Connes:1997cr}.
Among the numerous important developments in both frameworks, it is
worth noting two conjectures of utmost importance that signal the
developments in certain directions in string theory and not only,
related to the main theme of the present
review. The conjectures refer to
(i)~the duality among the 4-dimensional $N=4$ supersymmetric
Yang--Mills theory and the type IIB string theory on ${\rm AdS}_5 \times S^5$
\cite{Maldacena:1997re}; the former being the maximal $N=4$
supersymmetric Yang--Mills theory is known to
be UV all-loop f\/inite theory \cite{Mandelstam:1982cb,Brink:1982wv},
(ii)~the possibility of ``miraculous'' UV divergence cancellations in
4-dimensional maximal $N=8$ supergravity leading to a f\/inite theory,
as has been recently conf\/irmed in a remarkable 4-loop calculation
\cite{Bern:2009kd,Kallosh:2009jb,Bern:2007hh,Bern:2006kd,Green:2006yu}.
However, despite the importance of having frameworks to discuss
quantum gravity in a self-consistent way and possibly to construct
there f\/inite theories, it is very interesting to search for the
minimal realistic framework in which f\/initeness can take place.  In
addition, the main goal expected from a unif\/ied description of
interactions by the particle physics community is to understand the
present day large number of free parameters of the Standard Model (SM)
in terms of a few fundamental ones. In other words, to achieve {\it
  reduction of couplings} at a more fundamental level.

To reduce the number of free parameters of a theory, and thus render
it more predictive, one is usually led to introduce a symmetry.  Grand
Unif\/ied Theories (GUTs) are very good examples of such a procedure
\cite{Pati:1973rp,Georgi:1974sy,Georgi:1974yf,Fritzsch:1974nn,Carlson:1975gu}.
For instance, in the case of minimal $SU(5)$, because of (approximate)
gauge coupling unif\/ication, it was possible to reduce the gauge
couplings by one and give a prediction for one of them.
In fact, LEP data \cite{Amaldi:1991cn}
seem to
suggest that a further symmetry, namely $N=1$ global supersymmetry
\cite{Dimopoulos:1981zb,Sakai:1981gr}
should also be required to make the prediction viable.
GUTs can also
relate the Yukawa couplings among themselves, again $SU(5)$ provided
an example of this by predicting the ratio $M_{\tau}/M_b$
\cite{Buras:1977yy} in the SM.  Unfortunately, requi\-ring
more gauge symmetry does not seem to help, since additional
complications are introduced due to new degrees of freedom, in the
ways and channels of breaking the symmetry, and so on.

A natural extension of the GUT idea is to f\/ind a way to relate the
gauge and Yukawa sectors of a theory, that is to achieve gauge-Yukawa
Unif\/ication (GYU) \cite{Kubo:1995cg,Kubo:1997fi,Kobayashi:1999pn}.  A
symmetry which naturally relates the two sectors is supersymmetry, in
particular $N=2$ supersymmetry~\cite{Fayet:1978ig}.  It turns out, however, that $N=2$
supersymmetric theories have serious phenomenological problems due to
light mirror fermions.  Also in superstring theories and in composite
models there exist relations among the gauge and Yukawa couplings, but
both kind of theories have phenomenological problems, which we are not
going to address here.

There have been other attempts to relate the gauge and Yukawa sectors.
One was proposed by Decker, Pestieau, and Veltman
\cite{Decker:1979nk,Veltman:1980mj}.  By requiring the absence of
quadratic divergences in the SM, they found a relationship between the
squared masses appearing in the Yukawa and in the gauge sectors of the
theory.  A very similar relation is obtained by applying naively in
the SM the general formula derived from demanding spontaneous
supersymmetry breaking via F-terms \cite{Ferrara:1979wa}.  In both
cases a prediction for the top quark was possible only when it was
permitted experimentally to assume the $M_H \ll M_{W,Z}$ with the
result $M_t=69$~GeV.  Otherwise there is only a quadratic relation
among $M_t$ and $M_H$.  Using this relationship in the former case and
a version of {\em naturalness} into account, i.e.\ that the quadratic
corrections to the Higgs mass be at most equal to the physical mass,
the Higgs mass is found to be $\sim 260$~GeV, for a top quark mass of
around $176$ GeV \cite{Chaichian:1995ef}.  This value is already
excluded from the precision data \cite{Barate:2003sz}.

A well known relation among gauge and Yukawa couplings is the
Pendleton--Ross (P-R) {\em infrared} f\/ixed point
\cite{Pendleton:1980as}.  The P-R proposal, involving the Yukawa
coupling of the top quark $g_t$ and the strong gauge coupling
$\alpha_3$, was that the ratio $\alpha_t/\alpha_3$, where
$\alpha_t=g_t^2/4\pi$, has an infrared f\/ixed point.  This assumption
predicted $M_t \sim 100$ GeV.  In addition, it has been shown
\cite{Zimmermann:1992eg} that the P-R conjecture is not justif\/ied at
two-loops, since then the ratio $\alpha_t/ \alpha _3$ diverges in the
infrared.

Another interesting conjecture, made by Hill \cite{Hill:1980sq},
is that  $\alpha _t$ itself develops a quasi-infrared f\/ixed point,
leading to the prediction $M_t \sim 280$ GeV.

The P-R and Hill conjectures have been done in the framework of the SM.
The same conjectures within the Minimal Supersymmetric SM (MSSM) lead
to the following relations:
\begin{gather*}
M_t  \simeq   140~{\rm GeV}~\sin \beta \quad (\text{\rm P-R}),\qquad 
M_t  \simeq   200~{\rm GeV}~\sin \beta \quad (\text{\rm Hill}),
\end{gather*}
where $\tan \beta = {v_u/ v_d}$ is the ratio of the two VEV of the
Higgs f\/ields of the MSSM.
From theoretical considerations one can expect
\begin{gather*}
1 \lsim
\tan \beta \lsim 50 \quad \Leftrightarrow \quad
\frac{1}{\sqrt{2}} \lsim \sin\beta \lsim 1.
\end{gather*}
This corresponds to
\begin{gather*}
100~\mathrm{GeV}  \lsim M_t \lsim 140~\mathrm{GeV} \quad (\text{\rm P-R}),\qquad
140~\mathrm{GeV}  \lsim M_t \lsim 200~\mathrm{GeV} \quad (\text{\rm Hill}).
\end{gather*}
Thus, the MSSM P-R conjecture is ruled out, while within the MSSM, the
Hill conjecture does not give a prediction for $M_t$, since the value
of $\sin\beta$ is not f\/ixed by other considerations.
The Hill model can accommodate the correct value
 of $M_t \approx 173~\mathrm{GeV}$~\cite{:2009ec} for
$\sin\beta \approx 0.865$ corresponding to $\tan \beta \approx 1.7$.  Such
small values, however, are stongly challenged by the SUSY Higgs boson
searches at LEP~\cite{Schael:2006cr}. Only a very heavy scalar top
spectrum with large mixing could accommodate such a small $\tan \beta$ value.


In our studies
\cite{Kubo:1995cg,Kubo:1997fi,Kobayashi:1999pn,Kapetanakis:1992vx,Mondragon:1993tw,Kubo:1994bj,Kubo:1994xa,Kubo:1995zg,Kubo:1996js}
we have developed a complementary strategy in searching for a more
fundamental theory possibly at the Planck scale, whose basic
ingredients are GUTs and supersymmetry, but its consequences certainly
go beyond the known ones. Our method consists of hunting for
renormalization group invariant (RGI) relations holding below the
Planck scale, which in turn are preserved down to the GUT scale. This
programme, called gauge-Yukawa unif\/ication scheme, applied in the
dimensionless couplings of supersymmetric GUTs, such as gauge and
Yukawa couplings, had already noticable successes by predicting
correctly, among others, the top quark mass in the f\/inite and in the
minimal $N = 1$ supersymmetric $SU(5)$ GUTs
\cite{Kapetanakis:1992vx,Mondragon:1993tw,Kubo:1994bj}.
An impressive aspect of the RGI relations is that one can guarantee
their validity to all-orders in perturbation theory by studying the
uniqueness of the resulting relations at one-loop, as was proven in
the early days of the programme of {\it reduction of couplings}
\cite{Zimmermann:1984sx,Oehme:1984yy,Ma:1977hf,Ma:1984by}. Even more
remarkable is the fact that it is possible to f\/ind RGI relations among
couplings that guarantee f\/initeness to all-orders in perturbation
theory
\cite{Lucchesi:1987ef,Lucchesi:1987he,Lucchesi:1996ir,Ermushev:1986cu,Kazakov:1987vg}.

It is worth noting that the above principles have only  been applied in
supersymmetric GUTs for reasons that will be transparent in the
following sections.  We should also stress that our conjecture for GYU
is by no means in conf\/lict with the interesting proposals mentioned
before (see also~\cite{Schrempp:1992zi,Schrempp:1994xn,Schrempp:1996fb}), but it
rather uses all of them, hopefully in a more successful perspective.
For instance, the use of susy GUTs comprises the demand of the
cancellation of quadratic divergences in the SM.  Similarly, the very
interesting conjectures about the infrared f\/ixed points are
generalized in our proposal, since searching for RGI relations among
various couplings corresponds to searching for {\it fixed points} of
the coupled dif\/ferential equations obeyed by the various couplings of
a theory.

Although supersymmetry seems to be an essential feature for a
successful realization of the above programme, its breaking has to be
understood too, since it has the ambition to supply the SM with
predictions for several of its free parameters. Indeed, the search for
RGI relations has been extended to the soft supersymmetry breaking
sector (SSB) of these theories \cite{Kubo:1996js,Jack:1995gm}, which
involves parameters of dimension one and two.  Then a very interesting
progress has been made
\cite{Hisano:1997ua,Jack:1997pa,Avdeev:1997vx,Kazakov:1998uj,Kazakov:1997nf,Jack:1997eh,Kobayashi:1998jq}
concerning the renormalization properties of the SSB parameters based
conceptually and technically on the work of~\cite{Yamada:1994id}. In~\cite{Yamada:1994id} the powerful
supergraph method
\cite{Delbourgo:1974jg,Salam:1974pp,Fujikawa:1974ay,Grisaru:1979wc}
for studying supersymmetric theories has been applied to the softly
broken ones by using the ``spurion'' external space-time independent
superf\/ields \cite{Girardello:1981wz}.  In the latter method a~softly
broken supersymmetric gauge theory is considered as a supersymmetric
one in which the various parameters such as couplings and masses have
been promoted to external superf\/ields that acquire ``vacuum
expectation values''. Based on this method the relations among the
soft term renormalization and that of an unbroken supersymmetric
theory have been derived. In particular the $\beta$-functions of the
parameters of the softly broken theory are expressed in terms of
partial dif\/ferential operators involving the dimensionless parameters
of the unbroken theory. The key point in the strategy of~\cite{Kazakov:1998uj,Kazakov:1997nf,Jack:1997eh,Kobayashi:1998jq}
in solving the set of coupled dif\/ferential equations so as to be able
to express all parameters in a RGI way, was to transform the partial
dif\/ferential operators involved to total derivative operators.  This
is indeed possible to be done on the RGI surface which is def\/ined by
the solution of the reduction equations.

  On the phenomenological side there exist some serious developments
  too.  Previously an appealing ``universal'' set of soft scalar
  masses was asummed in the SSB sector of supersymmetric theories,
  given that apart from economy and simplicity (1)~they are part of
  the constraints that preserve f\/initeness up to two-loops
  \cite{Jones:1984cu,Jack:1994kd}, (2)~they are RGI up to two-loops in
  more general supersymmetric gauge theories, subject to the condition
  known as $P =1/3~Q$ \cite{Jack:1995gm} and (3)~they appear in the
  attractive dilaton dominated supersymmetry breaking superstring
  scenarios \cite{Ibanez:1992hc,Kaplunovsky:1993rd,Brignole:1993dj}.
  However, further studies have exhibited a number of problems all due
  to the restrictive nature of the ``universality'' assumption for the
  soft scalar masses.  For instance, (a)~in f\/inite unif\/ied theories the
  universality predicts that the lightest supersymmetric particle is a
  charged particle, namely the superpartner of the $\tau$ lepton
  $\tilde\tau$, (b)~the MSSM with universal soft scalar masses is
  inconsistent with the attractive radiative electroweak symmetry
  breaking~\cite{Brignole:1993dj}, and (c)~which is the worst of all,
  the universal soft scalar masses lead to charge and/or colour
  breaking minima deeper than the standard vacuum~\cite{Casas:1996wj}.
  Therefore, there have been attempts to relax this constraint without
  loosing its attractive features. First an interesting observation
  was made that in $N = 1$ gauge-Yukawa unif\/ied theories there exists
  a RGI sum rule for the soft scalar masses at lower orders; at
  one-loop for the non-f\/inite case \cite{Kawamura:1997cw} and at
  two-loops for the f\/inite case~\cite{Kobayashi:1997qx}. The sum rule
  manages to overcome the above unpleasant phenomenological
  consequences. Moreover it was proven~\cite{Kobayashi:1998jq} that
  the sum rule for the soft scalar massses is RGI to all-orders for
  both the general as well as for the f\/inite case. Finally, the exact
  $\beta$-function for the soft scalar masses in the
  Novikov--Shifman--Vainstein--Zakharov (NSVZ) scheme
  \cite{Novikov:1983ee,Novikov:1985rd,Shifman:1996iy} for the softly
  broken supersymmetric QCD has been obtained~\cite{Kobayashi:1998jq}.
  Armed with the above tools and results we are in a position to study
  the spectrum of the full f\/inite models in terms of few free
  parameters with emphasis on the predictions for the lightest Higgs
  mass, which is expected to be tested at LHC.

\section{Unif\/ication of couplings by the RGI method}

Let us next brief\/ly outline the idea of reduction of couplings.
Any RGI relation among couplings
(which does not depend on the renormalization
scale $\mu$ explicitly) can be expressed,
in the implicit form $\Phi (g_1,\dots,g_A) =\mbox{const}$,
which
has to satisfy the partial dif\/ferential equation (PDE)
\[
\mu \frac{d \Phi}{d \mu}  =  {\vec \nabla}\cdot {\vec \beta} =
\sum_{a=1}^{A}
 \beta_{a} \frac{\partial \Phi}{\partial g_{a}}=0,
\]
where $\beta_a$ is the $\beta$-function of $g_a$.
This PDE is equivalent
to a set of ordinary dif\/ferential equations,
the so-called reduction equations (REs) \cite{Zimmermann:1984sx,Oehme:1984yy,Oehme:1985jy},
\begin{gather}
\beta_{g}  \frac{d g_{a}}{d g} = \beta_{a}, \qquad a=1,\dots,A,
\label{redeq}
\end{gather}
where $g$ and $\beta_{g}$ are the primary
coupling and its $\beta$-function,
and the counting on $a$ does not include $g$.
Since maximally ($A-1$) independent
RGI ``constraints''
in the $A$-dimensional space of couplings
can be imposed by the $\Phi_a$'s, one could in principle
express all the couplings in terms of
a single coupling $g$.
 The strongest requirement is to demand
 power series solutions to the REs,
\begin{gather}
g_{a}  =  \sum_{n}\rho_{a}^{(n)} g^{2n+1},
\label{powerser}
\end{gather} which formally preserve perturbative renormalizability.
Remarkably, the uniqueness of such power series solutions can be
decided already at the one-loop level
\cite{Zimmermann:1984sx,Oehme:1984yy,Oehme:1985jy}.  To illustrate
this, let us assume that the $\beta$-functions have the form
\begin{gather*}
\beta_{a}  = \frac{1}{16 \pi^2}\left[ \sum_{b,c,d\neq
  g}\beta^{(1)\,bcd}_{a}g_b g_c g_d+
\sum_{b\neq g}\beta^{(1)\,b}_{a}g_b g^2\right]+\cdots, \\
\beta_{g}  = \frac{1}{16 \pi^2}\beta^{(1)}_{g}g^3+ \cdots,
\end{gather*} where
$\cdots$ stands for higher order terms, and $ \beta^{(1)\,bcd}_{a}$'s
are symmetric in $ b$, $c$, $d$.  We then assume that the $\rho_{a}^{(n)}$'s
with $n\leq r$ have been uniquely determined. To obtain
$\rho_{a}^{(r+1)}$'s, we insert the power series (\ref{powerser}) into
the REs (\ref{redeq}) and collect terms of ${\cal O}(g^{2r+3})$ and
f\/ind
\[
\sum_{d\neq g}M(r)_{a}^{d}\,\rho_{d}^{(r+1)}  =  \mbox{lower
  order quantities} ,
\]
where the r.h.s.\ is known by assumption,
and
\begin{gather*}
M(r)_{a}^{d}  = 3\sum_{b,c\neq
  g}\,\beta^{(1)\,bcd}_{a}\,\rho_{b}^{(1)}\,
\rho_{c}^{(1)}+\beta^{(1)\,d}_{a}
-(2r+1)\,\beta^{(1)}_{g}\,\delta_{a}^{d},\\ 
0  = \sum_{b,c,d\neq g}\,\beta^{(1)\,bcd}_{a}\,
\rho_{b}^{(1)}\,\rho_{c}^{(1)}\,\rho_{d}^{(1)} +\sum_{d\neq
  g}\beta^{(1)\,d}_{a}\,\rho_{d}^{(1)}
-\beta^{(1)}_{g}\,\rho_{a}^{(1)}.
\end{gather*}

 Therefore, the $\rho_{a}^{(n)}$'s for all $n > 1$ for a
given set of $\rho_{a}^{(1)}$'s can be uniquely determined if $\det
M(n)_{a}^{d} \neq 0$ for all $n \geq 0$.

As it will be clear later by examining specif\/ic examples,the various
couplings in supersymmetric theories have easily the same asymptotic
behaviour.  Therefore searching for a power series solution of the
form (\ref{powerser}) to the REs (\ref{redeq}) is justif\/ied. This is
not the case in non-supersymmetric theories, although the deeper
reason for this fact is not fully understood.

The possibility of coupling unif\/ication described in this section
is without any doubt
attractive because the ``completely reduced'' theory contains
only one independent coupling, but  it can be
unrealistic. Therefore, one often would like to impose fewer RGI
constraints, and this is the idea of partial reduction \cite{Kubo:1985up,Kubo:1988zu}.

\section{Reduction of dimensionful parameters}

The reduction of couplings
 was originally formulated for massless theories
on the basis of the Callan--Symanzik equation \cite{Zimmermann:1984sx,Oehme:1984yy,Oehme:1985jy}.
The extension to theories with massive parameters
is not straightforward if one wants to keep
the generality and the rigor
on the same level as for the massless case;
one has
to fulf\/ill a set of requirements coming from
the renormalization group
equations,  the  Callan--Symanzik equations, etc.
along with the normalization
conditions imposed on irreducible Green's functions \cite{Piguet:1989pc}.
See \cite{Zimmermann:2000hn} for  interesting results in this direction.
Here, to simplify the situation,  we would like  to
 assume  that
 a mass-independent renormalization scheme has been
employed so that all the  RG functions have only  trivial
dependencies of dimensional parameters.

To be general, we consider  a renormalizable theory
which contain a set of $(N+1)$ dimension-zero couplings,
$\{\hat{g}_0,\hat{g}_1,\dots,\hat{g}_N\}$, a~set of $L$
parameters with  dimension
one, $\{\hat{h}_1,\dots,\hat{h}_L\}$,
and a~set of $M$ parameters with dimension two,
$\{\hat{m}_{1}^{2},\dots,\hat{m}_{M}^{2}\}$.
The renormalized irreducible vertex function
satisf\/ies the RG equation
\begin{gather}
0  =  {\cal D}\Gamma \big[{\bf
\Phi}'s;\hat{g}_0,\hat{g}_1,\dots,\hat{g}_N;\hat{h}_1,\dots,\hat{h}_L;
\hat{m}^{2}_{1},\dots,\hat{m}^{2}_{M};\mu\big], \label{vertex}\\
{\cal D}  =  \mu\frac{\partial}{\partial \mu}+
 \sum_{i=0}^{N} \beta_i
\frac{\partial}{\partial \hat{g}_i}+
\sum_{a=1}^{L} \gamma_{a}^{h}
\frac{\partial}{\partial \hat{h}_a}+
\sum_{\alpha=1}^{M} \gamma^{m^2}_{\alpha}\frac{\partial}
{\partial \hat{m}_{\alpha}^{2}}+ \sum_{J} \Phi_I
\gamma^{\phi I}_{~~~J} \frac{\delta}{\delta \Phi_J}.\nonumber
\end{gather}
Since we assume a mass-independent renormalization scheme,
the $\gamma$'s have the form
\begin{gather*}
\gamma_{a}^{h}  =  \sum_{b=1}^{L}\,
\gamma_{a}^{h,b}(g_0,\dots,g_N)\hat{h}_b, \\
\gamma_{\alpha}^{m^2}  =
\sum_{\beta=1}^{M}\,\gamma_{\alpha}^{m^2,\beta}(g_0,\dots,g_N)
\hat{m}_{\beta}^{2}+
\sum_{a,b=1}^{L}\,\gamma_{\alpha}^{m^2,a b}
(g_0,\dots,g_N)\hat{h}_a \hat{h}_b,
\end{gather*}
where $\gamma_{a}^{h,b}$,
$\gamma_{\alpha}^{m^2,\beta}$ and
  $\gamma_{a}^{m^2,a b}$
are power series of the dimension-zero
couplings $g$'s in perturbation theory.

As in the massless case, we then look for
 conditions under which the reduction of
parameters,
\begin{gather}
\hat{g}_i  =  \hat{g}_i(g), \qquad i=1,\dots,N,\label{gr}\\
\hat{h}_a = \sum_{b=1}^{P}
f_{a}^{b}(g) h_b,\qquad a=P+1,\dots,L,\label{h}\\
\hat{m}_{\alpha}^{2} = \sum_{\beta=1}^{Q}
e_{\alpha}^{\beta}(g) m_{\beta}^{2}+
\sum_{a,b=1}^{P} k_{\alpha}^{a b}(g)
h_a h_b, \qquad \alpha=Q+1,\dots,M,\label{m}
\end{gather}
is consistent with the RG equation \eqref{vertex},
where we assume that $g\equiv g_0$, $h_a \equiv
\hat{h}_a$ $(1 \leq a \leq P)$ and
$m_{\alpha}^{2} \equiv
\hat{m}_{\alpha}^{2}$ $(1 \leq \alpha \leq Q)$
are independent parameters of the reduced theory.
We f\/ind  that the following set of
equations has to be satisf\/ied:
\begin{gather}
\beta_g \frac{\partial
\hat{g}_{i}}{\partial g}  = \beta_i , \qquad i=1,\dots,N, \label{betagr}\\
\beta_g \frac{\partial
\hat{h}_{a}}{\partial g}+\sum_{b=1}^{P}\gamma_{b}^{h}
\frac{\partial
\hat{h}_{a}}{\partial
h_b}  = \gamma_{a}^{h},\qquad a=P+1,\dots,L,\label{betah}\\
\beta_g \frac{\partial
\hat{m}^{2}_{\alpha}}{\partial g}
+\sum_{a=1}^{P}\gamma_{a}^{h}
\frac{\partial
\hat{m}^{2}_{\alpha}}{\partial
h_a}+
\sum_{\beta=1}^{Q}\gamma_{\beta}^{m^2}
\frac{\partial
\hat{m}^{2}_{\alpha}}{\partial
m^{2}_{\beta}}
  = \gamma_{\alpha}^{m^2}, \qquad \alpha=Q+1,\dots,M. \label{betam}
\end{gather}
Using equation (\ref{vertex}) for $\gamma$'s, one f\/inds that
equations~(\ref{betagr})--(\ref{betam})
reduce to
\begin{gather}
 \beta_g \frac{d f_{a}^{b}}{d g}+
\sum_{c=1}^{P}  f_{a}^{c}
\left[\gamma_{c}^{h,b}+\sum_{d=P+1}^{L}
\gamma_{c}^{h,d}f_{d}^{ b}\right]
-\gamma_{a}^{h,b}-\sum_{d=P+1}^{L}
\gamma_{a}^{h,d}f_{d}^{ b}=0,\label{red1}\\
\qquad a=P+1,\dots,L,\quad  b=1,\dots,P,\nonumber\\
 \beta_g \frac{d e_{\alpha}^{\beta}}{d g}+
\sum_{\gamma=1}^{Q}  e_{\alpha}^{\gamma}
\left[ \gamma_{\gamma}^{m^2,\beta}+\sum_{\delta=Q+1}^{M}
\gamma_{\gamma}^{m^2,\delta}e_{\delta}^{\beta} \right]
-\gamma_{\alpha}^{m^2,\beta}-\sum_{\delta=Q+1}^{M}
\gamma_{\alpha}^{m^2,\delta}e_{\delta}^{\beta} =0 ,\label{red2}\\
\qquad \alpha=Q+1,\dots,M,\quad  \beta=1,\dots,Q,\nonumber\\
 \beta_g \frac{d k_{\alpha}^{a b}}{d g}+2\sum_{c=1}^{P}
\left(\gamma_{c}^{h,a}+\sum_{d=P+1}^{L}
\gamma_{c}^{h,d}f_{d}^{a}\right)k_{\alpha}^{c b}+
\sum_{\beta=1}^{Q}\, e_{\alpha}^{\beta}
\left[\gamma_{\beta}^{m^2,a b}+\sum_{c,d=P+1}^{L}
\gamma_{\beta}^{m^2,c d}f_{c}^{a} f_{d}^{b}\right.\nonumber\\
\left. \qquad{} +2\sum_{c=P+1}^{L} \gamma_{\beta}^{m^2,c b}f_{c}^{a}+
\sum_{\delta=Q+1}^{M} \gamma_{\beta}^{m^2,\delta}
k_{\delta}^{a b} \right]-
\left[\gamma_{\alpha}^{m^2,a b}+\sum_{c,d=P+1}^{L}
\gamma_{\alpha}^{m^2,c d}f_{c}^{a} f_{d}^{b}\right.\nonumber\\
\left. \qquad{} +
2\sum_{c=P+1}^{L} \gamma_{\alpha}^{m^2,c b}f_{c}^{a}
+\sum_{\delta=Q+1}^{M} \gamma_{\alpha}^{m^2,\delta}
k_{\delta}^{a b} \right]=0,\label{red3}\\
\qquad \alpha=Q+1,\dots,M, \quad  a,b=1,\dots,P.\nonumber
\end{gather}
If these equations are satisf\/ied,
the irreducible vertex function of the reduced theory
\begin{gather*}
 \Gamma_R [ {\bf
\Phi}'s; g; h_1,\dots,h_P; m^{2}_{1},
\dots,\hat{m}^{2}_{Q};\mu]\nonumber\\
\qquad \equiv \Gamma [{\bf
\Phi}'s; g,\hat{g}_1(g),\dots,\hat{g}_N (g);
 h_1,\dots,h_P, \hat{h}_{P+1}(g,h),\dots,\hat{h}_L(g,h);\nonumber\\
\qquad \quad \; m^{2}_{1},\dots,\hat{m}^{2}_{Q},\hat{m}^{2}_{Q+1}(g,h,m^2),
\dots,\hat{m}^{2}_{M}(g,h,m^2);\mu]
\end{gather*}
has
the same renormalization group f\/low as the original one.

The requirement for the reduced theory to be perturbative
renormalizable means that the functions $\hat{g}_i $, $f_{a}^{b} $,
$e_{\alpha}^{\beta}$ and $k_{\alpha}^{a b}$, def\/ined in
equations (\ref{gr})--(\ref{m}), should have a power series expansion in the
primary coupling $g$:
\begin{gather*}
\hat{g}_{i}  =  g \sum_{n=0}^{\infty}
\rho_{i}^{(n)} g^{n} , \qquad
f_{a}^{b}= g\sum_{n=0}^{\infty} \eta_{a}^{b~(n)} g^{n},\nonumber\\
e_{\alpha}^{\beta} = \sum_{n=0}^{\infty} \xi_{\alpha}^{\beta~(n)}
g^{n}, \qquad  k_{\alpha}^{a b }= \sum_{n=0}^{\infty} \chi_{\alpha}^{a
  b~(n)} g^{n}.
\end{gather*}
To obtain the expansion coef\/f\/icients, we insert
the power series ansatz above into
equations~(\ref{betagr}), (\ref{red1})--(\ref{red3}) and require that the
equations are satisf\/ied at each order in $g$. Note that the existence
of a unique power series solution is a non-trivial matter: It depends
on the theory as well as on the choice of the set of independent
parameters.

\section[Finiteness in $N=1$ supersymmetric gauge theories]{Finiteness in $\boldsymbol{N=1}$ supersymmetric gauge theories}
\label{sec:futs}

Let us consider a chiral, anomaly free,
$N=1$ globally supersymmetric
gauge theory based on a~group $G$ with gauge coupling
constant $g$. The
superpotential of the theory is given by
\begin{gather}
W =  \frac{1}{2} m_{ij}  \phi_{i}\,\phi_{j}+
\frac{1}{6} C_{ijk}  \phi_{i} \phi_{j} \phi_{k},
\label{supot}
\end{gather}
where $m_{ij}$ and $C_{ijk}$ are gauge invariant tensors and
the matter f\/ield $\phi_{i}$ transforms
according to the irreducible representation  $R_{i}$
of the gauge group $G$. The
renormalization constants associated with the
superpotential (\ref{supot}), assuming that
supersymmetry is preserved, are
\begin{gather*}
\phi_{i}^{0} = (Z^{j}_{i})^{(1/2)} \phi_{j},\\
m_{ij}^{0} = Z^{i'j'}_{ij}\,m_{i'j'},\\
C_{ijk}^{0} = Z^{i'j'k'}_{ijk} C_{i'j'k'}.
\end{gather*}
The $N=1$ non-renormalization theorem \cite{Wess:1973kz,Iliopoulos:1974zv,Fujikawa:1974ay} ensures that
there are no mass
and cubic-interaction-term inf\/inities and therefore
\begin{gather*}
Z_{ijk}^{i'j'k'} Z^{1/2\,i''}_{i'} Z^{1/2\,j''}_{j'}
 Z^{1/2\,k''}_{k'} = \delta_{(i}^{i''}
 \delta_{j}^{j''}\delta_{k)}^{k''},\nonumber\\
Z_{ij}^{i'j'} Z^{1/2\,i''}_{i'} Z^{1/2\,j''}_{j'}
 = \delta_{(i}^{i''} \delta_{j)}^{j''}.
\end{gather*}
As a result the only surviving possible inf\/inities are
the wave-function renormalization constants~$Z^{j}_{i}$, i.e.,  one inf\/inity
for each f\/ield. The one-loop $\beta$-function of the gauge
coupling $g$ is given by~\cite{Parkes:1984dh}
\begin{gather}
\beta^{(1)}_{g}=\frac{d g}{d t} =
\frac{g^3}{16\pi^2}\left[\sum_{i} l(R_{i})-3\,C_{2}(G)\right],
\label{betag}
\end{gather}
where $l(R_{i})$ is the Dynkin index of $R_{i}$ and $C_{2}(G)$
 is the
quadratic Casimir of the adjoint representation of the
gauge group $G$. The $\beta$-functions of
$C_{ijk}$,
by virtue of the non-renormalization theorem, are related to the
anomalous dimension matrix $\gamma_{ij}$ of the matter f\/ields
$\phi_{i}$ as:
\begin{gather}
\beta_{ijk} =
 \frac{d C_{ijk}}{d t}=C_{ijl}\gamma^{l}_{k}+
 C_{ikl}\gamma^{l}_{j}+
 C_{jkl}\gamma^{l}_{i}.
\label{betay}
\end{gather}
At one-loop level $\gamma_{ij}$ is \cite{Parkes:1984dh}
\begin{gather}
\gamma^{i(1)}_j=\frac{1}{32\pi^2} \big[
C^{ikl} C_{jkl}-2 g^2 C_{2}(R_{i})\delta_{j}^1 \big],
\label{gamay}
\end{gather}
where $C_{2}(R_{i})$ is the quadratic Casimir of the representation
$R_{i}$, and $C^{ijk}=C_{ijk}^{*}$.
Since
dimensional coupling parameters such as masses  and couplings of cubic
scalar f\/ield terms do not inf\/luence the asymptotic properties
 of a theory on which we are interested here, it is
suf\/f\/icient to take into account only the dimensionless supersymmetric
couplings such as $g$ and $C_{ijk}$.
So we neglect the existence of dimensional parameters, and
assume furthermore that
$C_{ijk}$ are real so that $C_{ijk}^2$ always are positive numbers.

As one can see from equations~(\ref{betag}) and (\ref{gamay}),
 all the one-loop $\beta$-functions of the theory vanish if
 $\beta_g^{(1)}$ and $\gamma _{ij}^{(1)}$ vanish, i.e.
\begin{gather}
\sum _i \ell (R_i) = 3 C_2(G) ,
\label{1st}
\\
C^{ikl} C_{jkl} = 2\delta ^i_j g^2  C_2(R_i).
\label{2nd}
\end{gather}

The conditions for f\/initeness for $N=1$ f\/ield theories with $SU(N)$ gauge
symmetry are discussed in \cite{Rajpoot:1984zq}, and the
analysis of the anomaly-free and no-charge renormalization
requirements for these theories can be found in \cite{Rajpoot:1985aq}.
A very interesting result is that the conditions (\ref{1st}), (\ref{2nd})
are necessary and suf\/f\/icient for f\/initeness at the two-loop level
\cite{Parkes:1984dh,West:1984dg,Jones:1985ay,Jones:1984cx,Parkes:1985hh}.

In case supersymmetry is broken by soft terms, the requirement of
f\/initeness in the one-loop soft breaking terms imposes further
constraints among themselves \cite{Jones:1984cu}.  In addition, the same set
of conditions that are suf\/f\/icient for one-loop f\/initeness of the soft
breaking terms render the soft sector of the theory two-loop f\/inite
\cite{Jones:1984cu}.

The one- and two-loop f\/initeness conditions (\ref{1st}), (\ref{2nd}) restrict
considerably the possible choices of the irreps. $R_i$ for a given
group $G$ as well as the Yukawa couplings in the superpotential
(\ref{supot}).  Note in particular that the f\/initeness conditions cannot be
applied to the minimal supersymmetric standard model (MSSM), since the presence
of a $U(1)$ gauge group is incompatible with the condition~(\ref{1st}), due to $C_2[U(1)]=0$.  This naturally leads to the
expectation that f\/initeness should be attained at the grand unif\/ied
level only, the MSSM being just the corresponding, low-energy,
ef\/fective theory.

Another important consequence of one- and two-loop f\/initeness is that
supersymmetry (most probably) can only be broken due to the soft
breaking terms.  Indeed, due to the unacceptability of gauge singlets,
F-type spontaneous symmetry breaking \cite{O'Raifeartaigh:1975pr}
terms are incompatible with f\/initeness, as well as D-type
\cite{Fayet:1974jb} spontaneous breaking which requires the existence
of a $U(1)$ gauge group.

A natural question to ask is what happens at higher loop orders.  The
answer is contained in a theorem
\cite{Lucchesi:1987he,Lucchesi:1987ef} which states the necessary and
suf\/f\/icient conditions to achieve f\/initeness at all orders.  Before we
discuss the theorem let us make some introductory remarks.  The
f\/initeness conditions impose relations between gauge and Yukawa
couplings.  To require such relations which render the couplings
mutually dependent at a given renormalization point is trivial.  What
is not trivial is to guarantee that relations leading to a reduction
of the couplings hold at any renormalization point.  As we have seen,
the necessary and also suf\/f\/icient, condition for this to happen is to
require that such relations are solutions to the REs
\begin{gather}
\beta _g
\frac{d C_{ijk}}{dg} = \beta _{ijk}
\label{redeq2}
\end{gather}
and hold at all orders.   Remarkably, the existence of
all-order power series solutions to (\ref{redeq2}) can be decided at
one-loop level, as already mentioned.

Let us now turn to the all-order f\/initeness theorem
\cite{Lucchesi:1987he,Lucchesi:1987ef}, which states that if a $N=1$
supersymmetric gauge theory can become f\/inite to all orders in the
sense of vanishing $\beta$-functions, that is of physical scale
invariance.  It is based on (a)~the structure of the supercurrent in
$N=1$ supersymmetric gauge theory
\cite{Ferrara:1974pz,Piguet:1981mu,Piguet:1981mw}, and on (b)~the
non-renormalization properties of $N=1$ chiral anomalies
\cite{Lucchesi:1987he,Lucchesi:1987ef,Piguet:1986td,Piguet:1986pk,Ensign:1987wy}.
Details on the proof can be found in~\cite{Lucchesi:1987he,Lucchesi:1987ef} and further discussion in~\cite{Piguet:1986td,Piguet:1986pk,Ensign:1987wy,Lucchesi:1996ir,Piguet:1996mx}.
Here, following mostly~\cite{Piguet:1996mx} we present a
comprehensible sketch of the proof.

Consider a $N=1$ supersymmetric gauge theory, with simple Lie group
$G$.  The content of this theory is given at the classical level by
the matter supermultiplets $S_i$, which contain a scalar f\/ield
$\phi_i$ and a Weyl spinor $\psi_{ia}$, and the vector supermultiplet
$V_a$, which contains a gauge vector f\/ield $A_{\mu}^a$ and a~gaugino
Weyl spinor $\lambda^a_{\alpha}$.

Let us f\/irst recall certain facts about the theory:

(1)~A massless $N=1$ supersymmetric theory is invariant
under a $U(1)$ chiral transformation~$R$ under which the various f\/ields
transform as follows
\begin{gather*}
A'_{\mu} = A_{\mu},\qquad \lambda '_{\alpha}=\exp({-i\theta})\lambda_{\alpha},\\
\phi ' =  \exp\left({-i\frac{2}{3}\theta}\right)\phi,\qquad \psi_{\alpha}'= \exp\left({-i\frac{1}
    {3}\theta}\right)\psi_{\alpha}, \qquad \dots.
\end{gather*}
The corresponding axial Noether current $J^{\mu}_R(x)$ is
\begin{gather}
J^{\mu}_R(x)=\bar{\lambda}\gamma^{\mu}\gamma^5\lambda + \cdots
\label{noethcurr}
\end{gather}
is conserved classically, while in the quantum case is violated by the
axial anomaly
\begin{gather}
\partial_{\mu} J^{\mu}_R =
r(\epsilon^{\mu\nu\sigma\rho}F_{\mu\nu}F_{\sigma\rho}+\cdots).
\label{anomaly}
\end{gather}

From its known topological origin in ordinary gauge theories
\cite{AlvarezGaume:1983cs,Bardeen:1984pm,Zumino:1983rz}, one would expect that the axial vector current
$J^{\mu}_R$ to satisfy the Adler--Bardeen theorem  and
receive corrections only at the one-loop level.  Indeed it has been
shown that the same non-renormalization theorem holds also in
supersymmetric theories \cite{Piguet:1986td,Piguet:1986pk,Ensign:1987wy}.  Therefore
\begin{gather}
r=\hbar \beta_g^{(1)}.
\label{r}
\end{gather}

(2)~The massless theory we consider is scale invariant at
the classical level and, in general, there is a scale anomaly due to
radiative corrections.  The scale anomaly appears in the trace of the
energy momentum tensor $T_{\mu\nu}$, which is traceless classically.
It has the form
\begin{gather}
T^{\mu}_{\mu} = \beta_g F^{\mu\nu}F_{\mu\nu} +\cdots.
\label{Tmm}
\end{gather}

 (3)  Massless, $N=1$ supersymmetric gauge theories are
classically invariant under the supersymmetric extension of the
conformal group~-- the superconformal group.  Examining the
superconformal algebra, it can be seen that the subset of
superconformal transformations consisting of translations,
supersymmetry transformations, and axial $R$ transformations is closed
under supersymmetry, i.e.\ these transformations form a representation
of supersymmetry.  It follows that the conserved currents
corresponding to these transformations make up a supermultiplet
represented by an axial vector superf\/ield called supercurrent~$J$,
\begin{gather}
J \equiv \{ J'^{\mu}_R, Q^{\mu}_{\alpha}, T^{\mu}_{\nu} , \dots \},
\label{J}
\end{gather}
where $J'^{\mu}_R$ is the current associated to $R$ invariance,
$Q^{\mu}_{\alpha}$ is the one associated to supersymmetry invariance,
and $T^{\mu}_{\nu}$ the one associated to translational invariance
(energy-momentum tensor).

The anomalies of the $R$ current $J'^{\mu}_R$, the trace
anomalies of the
supersymmetry current, and the energy-momentum tensor, form also
a second supermultiplet, called the supertrace anomaly
\begin{gather*}
S = \{{\rm Re}\, S, {\rm Im}\, S, S_{\alpha}\} =  \big\{T^{\mu}_{\mu},~\partial _{\mu} J'^{\mu}_R,~\sigma^{\mu}_{\alpha
  \dot{\beta}} \bar{Q}^{\dot\beta}_{\mu}+\cdots \big\},
\end{gather*}
where $T^{\mu}_{\mu}$ in equation~(\ref{Tmm}) and
\begin{gather*}
\partial _{\mu} J'^{\mu}_R  = \beta_g\epsilon^{\mu\nu\sigma\rho}
F_{\mu\nu}F_{\sigma\rho}+\cdots, \qquad
\sigma^{\mu}_{\alpha \dot{\beta}} \bar{Q}^{\dot\beta}_{\mu} = \beta_g
\lambda^{\beta}\sigma^{\mu\nu}_{\alpha\beta}F_{\mu\nu}+\cdots.
\end{gather*}

 (4) It is very important to note that
the Noether current def\/ined in (\ref{noethcurr}) is not the same as the
current associated to $R$ invariance that appears in the
supercurrent
$J$ in (\ref{J}), but they coincide in the tree approximation.
So starting from a unique classical Noether current
$J^{\mu}_{R({\rm class})}$,  the Noether
current $J^{\mu}_R$ is def\/ined as the quantum extension of
$J^{\mu}_{R({\rm class})}$ which allows for the
validity of the non-renormalization theorem.  On the other hand
$J'^{\mu}_R$, is def\/ined to belong to the supercurrent $J$,
together with the energy-momentum tensor.  The two requirements
cannot be fulf\/illed by a single current operator at the same time.

Although the Noether current $J^{\mu}_R$ which obeys (\ref{anomaly})
and the current $J'^{\mu}_R$ belonging to the supercurrent multiplet
$J$ are not the same, there is a relation
\cite{Lucchesi:1987he,Lucchesi:1987ef} between quantities associated
with them
\begin{gather}
r=\beta_g(1+x_g)+\beta_{ijk}x^{ijk}-\gamma_Ar^A,
\label{rbeta}
\end{gather}
where $r$ was given in equation~(\ref{r}).  The $r^A$ are the
non-renormalized coef\/f\/icients of
the anomalies of the Noether currents associated to the chiral
invariances of the superpotential, and~-- like~$r$~-- are strictly
one-loop quantities. The $\gamma_A$'s are linear
combinations of the anomalous dimensions of the matter f\/ields, and
$x_g$, and $x^{ijk}$ are radiative correction quantities.
The structure of equality (\ref{rbeta}) is independent of the
renormalization scheme.

One-loop f\/initeness, i.e.\ vanishing of the $\beta$-functions at one-loop,
implies that the Yukawa couplings $\lambda_{ijk}$ must be functions of
the gauge coupling~$g$. To f\/ind a similar condition to all orders it
is necessary and suf\/f\/icient for the Yukawa couplings to be a formal
power series in~$g$, which is solution of the REs~(\ref{redeq2}).

We can now state the theorem for all-order vanishing
$\beta$-functions.

\begin{theorem}
Consider an $N=1$ supersymmetric Yang--Mills theory, with simple gauge
group. If the following conditions are satisfied
\begin{enumerate}\itemsep=0pt
\item[$1)$] there is no gauge anomaly;
\item[$2)$] the gauge $\beta$-function vanishes at one-loop
  \[
  \beta^{(1)}_g = 0 =\sum_i l(R_{i})-3 C_{2}(G);
  \]
\item[$3)$] There exist solutions of the form
  \begin{gather}
  C_{ijk}=\rho_{ijk}g, \qquad \rho_{ijk}\in{\mathbb C}
  \label{soltheo}
  \end{gather}
to the  conditions of vanishing one-loop matter fields anomalous dimensions
  \begin{gather*}
   \gamma^{i~(1)}_j = 0\\
   \phantom{\gamma^{i~(1)}_j}{} =\frac{1}{32\pi^2} \big[
  C^{ikl} C_{jkl}-2 g^2 C_{2}(R_{i})\delta_{ij} \big];
  \end{gather*}
\item[$4)$] these solutions are isolated and non-degenerate when considered
  as solutions of vanishing one-loop Yukawa $\beta$-functions:
   \[
   \beta_{ijk}=0.
   \]
\end{enumerate}
Then, each of the solutions \eqref{soltheo} can be uniquely extended
to a formal power series in~$g$, and the associated super Yang--Mills
models depend on the single coupling constant~$g$ with a $\beta$-function which vanishes at all-orders.
\end{theorem}

It is important to note a few things:
The requirement of isolated and non-degenerate
solutions guarantees the
existence of a unique formal power series solution to the reduction
equations.
The vanishing of the gauge $\beta$-function at one-loop,
$\beta_g^{(1)}$, is equivalent to the
vanishing of the $R$ current anomaly (\ref{anomaly}).  The vanishing of
the anomalous
dimensions at one-loop implies the vanishing of the Yukawa couplings
$\beta$-functions at that order.  It also implies the vanishing of the
chiral anomaly coef\/f\/icients $r^A$.  This last property is a necessary
condition for having $\beta$-functions vanishing at all orders\footnote{There is an alternative way to f\/ind f\/inite theories~\cite{Leigh:1995ep}.}.

\begin{proof}
Insert $\beta_{ijk}$ as given by the REs into the
relationship (\ref{rbeta}) between the axial anomalies coef\/f\/icients and
the $\beta$-functions.  Since these chiral anomalies vanish, we get
for $\beta_g$ an homogeneous equation of the form
\begin{gather*}
0=\beta_g(1+O(\hbar)).
\end{gather*}
The solution of this equation in the sense of a formal power series in
$\hbar$ is $\beta_g=0$, order by order.  Therefore, due to the
REs (\ref{redeq2}), $\beta_{ijk}=0$ too.
\end{proof}

Thus we see that f\/initeness and reduction of couplings are intimately
related. Since an equation like equation~(\ref{rbeta}) is lacking in
non-supersymmetric theories, one cannot extend the validity of a
similar theorem in such theories.

\section[Sum rule for SB terms in $N=1$ supersymmetric and finite theories: all-loop results]{Sum rule for SB terms in $\boldsymbol{N=1}$ supersymmetric\\ and f\/inite theories: all-loop results}

The method of reducing the dimensionless couplings has been
extended~\cite{Kubo:1996js}, to the soft supersymmetry breaking (SSB)
dimensionful parameters of $N = 1$ supersymmetric theories.  In
addition it was found \cite{Kawamura:1997cw} that RGI SSB scalar
masses in gauge-Yukawa unif\/ied models satisfy a universal sum rule.
Here we will describe f\/irst how the use of the available two-loop RG
functions and the requirement of f\/initeness of the SSB parameters up
to this order leads to the soft scalar-mass sum rule~\cite{Kobayashi:1997qx}.

Consider the superpotential given by (\ref{supot})
along with the Lagrangian for SSB terms
\begin{gather*}
-{\cal L}_{\rm SB}  =
\frac{1}{6}  h^{ijk} \phi_i \phi_j \phi_k
+
\frac{1}{2}  b^{ij} \phi_i \phi_j  +
\frac{1}{2}  \big(m^2\big)^{j}_{i} \phi^{*\,i} \phi_j+
\frac{1}{2}  M \lambda \lambda+\mbox{h.c.},
\end{gather*}
where the $\phi_i$ are the
scalar parts of the chiral superf\/ields $\Phi_i$, $\lambda$ are the gauginos
and $M$ their unif\/ied mass.
Since we would like to consider
only f\/inite theories here, we assume that
the gauge group is  a simple group and the one-loop
$\beta$-function of the
gauge coupling $g$  vanishes.
We also assume that the reduction equations
admit power series solutions of the form
\[
C^{ijk}  =  g \sum_{n} \rho^{ijk}_{(n)} g^{2n}.
\]
According to the f\/initeness theorem
of~\cite{Lucchesi:1987ef,Lucchesi:1987he}, the theory is then f\/inite to all orders in
perturbation theory, if, among others, the one-loop anomalous dimensions
$\gamma_{i}^{j(1)}$ vanish.  The one- and two-loop f\/initeness for
$h^{ijk}$ can be achieved by \cite{Jack:1994kd}
\begin{gather}
 h^{ijk}  =  -M C^{ijk}+\cdots =-M
\rho^{ijk}_{(0)} g+O\big(g^5\big) ,
\label{hY}
\end{gather}
where $\cdots$ stand for  higher order terms.

Now, to obtain the two-loop sum rule for
soft scalar masses, we assume that
the lowest order coef\/f\/icients $\rho^{ijk}_{(0)}$
and also $(m^2)^{i}_{j}$ satisfy the diagonality relations
\begin{gather*}
\rho_{ipq(0)}\rho^{jpq}_{(0)}  \propto   \delta_{i}^{j} \quad \mbox{for all}
 \ p  \ \mbox{and} \ q \quad \mbox{and} \quad
\big(m^2\big)^{i}_{j}= m^{2}_{j}\delta^{i}_{j},
\end{gather*}
respectively.
Then we f\/ind the following soft scalar-mass sum
rule \cite{Kobayashi:1997qx,Kobayashi:1999pn,Mondragon:2003bp}
\begin{gather}
\big(m_{i}^{2}+m_{j}^{2}+m_{k}^{2}\big)/
M M^{\dag}  =
1+\frac{g^2}{16 \pi^2} \Delta^{(2)}+O\big(g^4\big)
\label{sumr}
\end{gather}
for $i$, $j$, $k$ with $\rho^{ijk}_{(0)} \neq 0$, where $\Delta^{(2)}$ is
the two-loop correction
\begin{gather*}
\Delta^{(2)} =   -2\sum_{l} \big[\big(m^{2}_{l}/M M^{\dag}\big)-(1/3)\big] T(R_l),
\end{gather*}
which vanishes for the
universal choice in accordance with the previous f\/indings~\cite{Jack:1994kd}.

If we know higher-loop $\beta$-functions explicitly, we can follow the same
procedure and f\/ind higher-loop RGI relations among SSB terms.
However, the $\beta$-functions of the soft scalar masses are explicitly
known only up to two loops.
In order to obtain higher-loop results some relations among
$\beta$-functions are needed.

Making use of the spurion technique
\cite{Delbourgo:1974jg,Salam:1974pp,Fujikawa:1974ay,Grisaru:1979wc,Girardello:1981wz}, it is possible to f\/ind
the following  all-loop relations among SSB $\beta$-functions,
\cite{Hisano:1997ua,Jack:1997pa,Avdeev:1997vx,Kazakov:1998uj,Kazakov:1997nf,Jack:1997eh}
\begin{gather*}
\beta_M  =  2{\cal O}\left(\frac{\beta_g}{g}\right) ,\\
\beta_h^{ijk} = \gamma^i{}_lh^{ljk}+\gamma^j{}_lh^{ilk}
+\gamma^k{}_lh^{ijl} -2\gamma_1^i{}_lC^{ljk}
-2\gamma_1^j{}_lC^{ilk}-2\gamma_1^k{}_lC^{ijl},\\
(\beta_{m^2})^i{}_j =\left[ \Delta
+ X \frac{\partial}{\partial g}\right]\gamma^i{}_j,\\
{\cal O} =\left(Mg^2\frac{\partial}{\partial g^2}
-h^{lmn}\frac{\partial}{\partial C^{lmn}}\right),\\
\Delta = 2{\cal O}{\cal O}^* +2|M|^2 g^2\frac{\partial}
{\partial g^2} +\tilde{C}_{lmn}
\frac{\partial}{\partial C_{lmn}} +
\tilde{C}^{lmn}\frac{\partial}{\partial C^{lmn}},
\end{gather*}
where $(\gamma_1)^i{}_j={\cal O}\gamma^i{}_j$,
$C_{lmn} = (C^{lmn})^*$, and
\begin{gather*}
\tilde{C}^{ijk} =
\big(m^2\big)^i{}_lC^{ljk}+\big(m^2\big)^j{}_lC^{ilk}+\big(m^2\big)^k{}_lC^{ijl}.
\end{gather*}
It was also found \cite{Jack:1997pa}  that the relation
\begin{gather*}
h^{ijk}  =  -M (C^{ijk})'
\equiv -M \frac{d C^{ijk}(g)}{d \ln g},
\end{gather*}
among couplings is all-loop RGI. Furthermore, using the all-loop gauge
$\beta$-function of Novikov  et al.~\cite{Novikov:1983ee,Novikov:1985rd,Shifman:1996iy} given
by
\begin{gather*}
\beta_g^{\rm NSVZ}  =
\frac{g^3}{16\pi^2}
\left[ \frac{\sum_l T(R_l)(1-\gamma_l /2)
-3 C(G)}{ 1-g^2C(G)/8\pi^2}\right],
\end{gather*}
it was found the all-loop RGI sum rule \cite{Kobayashi:1998jq},
\begin{gather}
m^2_i+m^2_j+m^2_k  =
|M|^2 \left\{
\frac{1}{1-g^2 C(G)/(8\pi^2)}\frac{d \ln C^{ijk}}{d \ln g}
+\frac{1}{2}\frac{d^2 \ln C^{ijk}}{d (\ln g)^2} \right\}\nonumber\\
 \phantom{m^2_i+m^2_j+m^2_k  =}{} +\sum_l
\frac{m^2_l T(R_l)}{C(G)-8\pi^2/g^2}
\frac{d \ln C^{ijk}}{d \ln g}.
\label{sum2}
\end{gather}
In addition
the exact-$\beta$-function for $m^2$
in the NSVZ scheme has been obtained \cite{Kobayashi:1998jq} for the f\/irst time and
is given by
\begin{gather*}
\beta_{m^2_i}^{\rm NSVZ}  = \left[
|M|^2 \left\{
\frac{1}{1-g^2 C(G)/(8\pi^2)}\frac{d }{d \ln g}
+\frac{1}{2}\frac{d^2 }{d (\ln g)^2}\right\}\right.\nonumber\\
 \left. \phantom{\beta_{m^2_i}^{\rm NSVZ}  =}{} +\sum_l
\frac{m^2_l T(R_l)}{C(G)-8\pi^2/g^2}
\frac{d }{d \ln g} \right] \gamma_{i}^{\rm NSVZ}.
\end{gather*}
Surprisingly enough, the all-loop result (\ref{sum2}) coincides with
the superstring result for the f\/inite case in a certain class of
orbifold models \cite{Kobayashi:1997qx} if
$d \ln C^{ijk}/{d \ln g}=1$.


\section[Finite ${SU(5)}$ Unified Theories]{Finite $\boldsymbol{SU(5)}$ Unif\/ied Theories}

Finite Unif\/ied Theories (FUTs) have always attracted interest for
their intriguing mathematical properties and their predictive power.
One very important result is that the one-loop f\/initeness conditions
(\ref{betay}), (\ref{gamay}) are suf\/f\/icient to guarantee two-loop
f\/initeness \cite{Parkes:1984dh}.  A classif\/ication of possible
one-loop f\/inite models was done by two groups
\cite{Hamidi:1984ft,Jiang:1988na,Jiang:1987hv}.  The f\/irst one and
two-loop f\/inite $SU(5)$ model was presented in~\cite{Jones:1984qd},
and shortly afterwards the conditions for f\/initeness in the soft
SUSY-breaking sector at one-loop~\cite{Jones:1984cx} were given.  In
\cite{Leon:1985jm} a one and two-loop f\/inite $SU(5)$ model was
presented where the rotation of the Higgs sector was proposed as a way
of making it realistic.  The f\/irst all-loop f\/inite theory was studied
in~\cite{Kapetanakis:1992vx,Mondragon:1993tw}, without taking into
account the soft breaking terms. Finite soft breaking terms and the
proof that one-loop f\/initeness in the soft terms also implies two-loop
f\/initeness was done in~\cite{Jack:1994kd}.  The inclusion of soft
breaking terms in a realistic model was done in~\cite{Kazakov:1995cy}
and their f\/initeness to all-loops studied in~\cite{Kazakov:1997nf},
although the universality of the soft breaking terms lead to a charged
LSP. This fact was also noticed in \cite{Yoshioka:1997yt}, where the
inclusion of an extra parameter in the boundary condition of the Higgs
mixing mass parameter was introduced to alleviate
it.
The derivation of the sum-rule in the soft supersymmetry breaking
sector and the proof that it can be made all-loop f\/inite were done in~\cite{Kobayashi:1997qx} and \cite{Kobayashi:1998jq}
respectively, allowing thus for the construction of all-loop f\/inite
realistic models.

From the classif\/ication of theories with vanishing one-loop gauge
$\beta$-function \cite{Hamidi:1984ft}, one can easily see that there
exist only two candidate possibilities to construct $SU(5)$ GUTs with
three generations. These possibilities require that the theory should
contain as matter f\/ields the chiral supermultiplets ${\bf
  5}$, $\overline{\bf 5}$, ${\bf 10}$, $\overline{\bf 5}$, ${\bf 24}$ with
the multiplicities $(6,9,4,1,0)$ and $(4,7,3,0,1)$, respectively. Only
the second one contains a ${\bf 24}$-plet which can be used to provide
the spontaneous symmetry breaking (SB) of $SU(5)$ down to $SU(3)\times
SU(2)\times U(1)$. For the f\/irst model one has to incorporate another
way, such as the Wilson f\/lux breaking mechanism to achieve the desired
SB of $SU(5)$ \cite{Kapetanakis:1992vx,Mondragon:1993tw}. Therefore,
for a self-consistent f\/ield theory discussion we would like to
concentrate only on the second possibility.

The particle content of the models we will study consists of the
following supermultiplets: three ($\overline{\bf 5} + \bf{10}$),
needed for each of the three generations of quarks and leptons, four
($\overline{\bf 5} + {\bf 5}$) and one ${\bf 24}$ considered as Higgs
supermultiplets.
When the gauge group of the f\/inite GUT is broken the theory is no
longer f\/inite, and we will assume that we are left with the MSSM.

Therefore, a predictive gauge-Yukawa unif\/ied $SU(5)$
model which is f\/inite to all orders, in addition to the requirements
mentioned already, should also have the following properties:

\begin{enumerate}\itemsep=0pt

\item
One-loop anomalous dimensions are diagonal,
i.e.,  $\gamma_{i}^{(1)\,j} \propto \delta^{j}_{i} $.
\item The three fermion generations, in the irreducible representations
  $\overline{\bf 5}_{i}$, ${\bf 10}_i$ $(i=1,2,3)$,  should
  not couple to the adjoint ${\bf 24}$.
\item The two Higgs doublets of the MSSM should mostly be made out of a
pair of Higgs quintet and anti-quintet, which couple to the third
generation.
\end{enumerate}

In the following we discuss two versions of the all-order f\/inite
model.  The model~\cite{Kapetanakis:1992vx,Mondragon:1993tw},
which will be labeled ${\bf A}$, and a slight variation of this model
(labeled~${\bf B}$), which can also be obtained from the class of the
models suggested in~\cite{Avdeev:1997vx,Kazakov:1998uj} with a
modif\/ication to suppress non-diagonal anomalous dimensions~\cite{Kobayashi:1997qx}.

The superpotential which describes the two models before the reduction
of couplings takes places is of the form
\cite{Kapetanakis:1992vx,Mondragon:1993tw,Kobayashi:1997qx,Jones:1984qd,Leon:1985jm}
\begin{gather}
W  =
\sum_{i=1}^{3} \left[ \frac{1}{2}g_{i}^{u}  {\bf 10}_i{\bf 10}_i H_{i}+
g_{i}^{d} {\bf 10}_i \overline{\bf 5}_{i}  \overline{H}_{i}\right] +
g_{23}^{u}\ {\bf 10}_2{\bf 10}_3 H_{4}\nonumber \\
\phantom{W=}{} +g_{23}^{d} {\bf 10}_2 \overline{\bf 5}_{3}  \overline{H}_{4}+
g_{32}^{d} {\bf 10}_3 \overline{\bf 5}_{2}  \overline{H}_{4}+
\sum_{a=1}^{4}g_{a}^{f} H_{a}  {\bf 24} \overline{H}_{a}+
\frac{g^{\lambda}}{3} ({\bf 24})^3,\label{zoup-super1}
\end{gather}
where
$H_{a}$ and $\overline{H}_{a}$ $(a=1,\dots,4)$
stand for the Higgs quintets and anti-quintets.

 The main dif\/ference between model ${\bf A}$ and model
${\bf B}$ is that two pairs of Higgs quintets and anti-quintets couple
to the ${\bf 24}$ in ${\bf B}$, so that it is not necessary to mix
them with $H_{4}$ and $\overline{H}_{4}$ in order to achieve the
triplet-doublet splitting after the symmetry breaking of $SU(5)$
\cite{Kobayashi:1997qx}.  Thus, although the particle content is the
same, the solutions to equations (\ref{betay}), (\ref{gamay}) and the sum rules
are dif\/ferent, which will ref\/lect in the phenomenology, as we will
see.

\subsection{FUTA}
\begin{table}[t]
\centering

\caption{Charges of the $Z_7\times Z_3\times Z_2$ symmetry for model
    \FUTA.}\label{tableA}
    \vspace{1mm}

\renewcommand{\arraystretch}{1.3}
\begin{tabular}{|l|l|l|l|l|l|l|l|l|l|l|l|l|l|l|l|}
\hline
& $\overline{{\bf 5}}_{1} $ & $\overline{{\bf 5}}_{2} $& $\overline{{\bf
    5}}_{3}$ & ${\bf 10}_{1} $ &  ${\bf 10}_{2}$ & ${\bf
  10}_{3} $ & $H_{1} $ & $H_{2} $ & $H_{3} $ &$H_{4 }$&  $\overline H_{1} $ &
$\overline H_{2} $ & $\overline H_{3} $ &$\overline H_{4 }$& ${\bf 24} $\\ \hline
$Z_7$ & 4 & 1 & 2 & 1 & 2 & 4 & 5 & 3 & 6 & $-5$ & $-3$ & $-6$ &0& 0 & 0 \\\hline
$Z_3$ & 0 & 0 & 0 & 1 & 2 & 0 & 1 & 2 & 0 & $-1$ & $-2$ & 0 & 0 & 0&0  \\\hline
$Z_2$ & 1 & 1 & 1 & 1 & 1 & 1 & 0 & 0 & 0 & 0 & 0 & 0 &  0 & 0 &0 \\\hline
\end{tabular}

\renewcommand{\arraystretch}{1.0}

\end{table}

After the reduction of couplings
the symmetry of the superpotential $W$ (\ref{zoup-super1}) is enhanced.
For  model ${\bf A}$ one f\/inds that
the superpotential has the
$Z_7\times Z_3\times Z_2$ discrete symmetry with the charge assignment
as shown in Table \ref{tableA}, and with the following superpotential
\begin{gather*}
W_A = \sum_{i=1}^{3} \left[ \frac{1}{2}g_{i}^{u}
 {\bf 10}_i{\bf 10}_i H_{i}+
g_{i}^{d} {\bf 10}_i \overline{\bf 5}_{i}
\overline{H}_{i}\right] +
g_{4}^{f} H_{4}
{\bf 24} \overline{H}_{4}+
\frac{g^{\lambda}}{3} ({\bf 24})^3.
\end{gather*}

The non-degenerate and isolated solutions to $\gamma^{(1)}_{i}=0$ for
 model \FUTA, which are the boundary conditions for the Yukawa
 couplings at the GUT scale, are:
\begin{gather}
  (g_{1}^{u})^2
=\frac{8}{5} g^2, \qquad \big(g_{1}^{d}\big)^2
=\frac{6}{5} g^2 ,\qquad
(g_{2}^{u})^2=(g_{3}^{u})^2=\frac{8}{5} g^2 ,\nonumber\\
  \big(g_{2}^{d}\big)^2 = \big(g_{3}^{d}\big)^2=\frac{6}{5} g^2,\qquad
(g_{23}^{u})^2 =0 ,\qquad
\big(g_{23}^{d}\big)^2=\big(g_{32}^{d}\big)^2=0,
\nonumber\\
  \big(g^{\lambda}\big)^2 =\frac{15}{7}g^2 ,\qquad  \big(g_{2}^{f}\big)^2
=\big(g_{3}^{f}\big)^2=0 ,\qquad  \big(g_{1}^{f}\big)^2=0 ,\qquad
\big(g_{4}^{f}\big)^2= g^2.\label{zoup-SOL5}
\end{gather}
In the dimensionful sector, the sum rule gives us the following
boundary conditions at the GUT scale for this model
\cite{Kobayashi:1997qx}:
\[
m^{2}_{H_u}+
2  m^{2}_{{\bf 10}}  =
m^{2}_{H_d}+ m^{2}_{\overline{{\bf 5}}}+
m^{2}_{{\bf 10}}=M^2  ,
\]
and thus we are left with only three free parameters, namely
$m_{\overline{{\bf 5}}}\equiv m_{\overline{{\bf 5}}_3}$,
$m_{{\bf 10}}\equiv m_{{\bf 10}_3}$
and~$M$.

\subsection{FUTB}

\begin{table}[t]
\centering   \renewcommand{\arraystretch}{1.3}

\caption{Charges of the $Z_4\times Z_4\times Z_4$ symmetry for model     \FUTB.}\label{tableB}

\vspace{1mm}

\begin{tabular}{|l|l|l|l|l|l|l|l|l|l|l|l|l|l|l|l|}
\hline
& $\overline{{\bf 5}}_{1} $ & $\overline{{\bf 5}}_{2} $& $\overline{{\bf
    5}}_{3}$ & ${\bf 10}_{1} $ &  ${\bf 10}_{2}$ &  ${\bf
  10}_{3} $ & $ H_{1} $ & $H_{2} $ & $ H_{3}
$ &$H_{4}$&   $\overline H_{1} $ &
$\overline H_{2} $ & $\overline H_{3} $ &$\overline H_{4 }$&${\bf 24} $\\\hline
$Z_4$ & 1 & 0 & 0 & 1 & 0 & 0 & 2 & 0 & 0 & 0 & $-2$ & 0 & 0 & 0 &0  \\\hline
$Z_4$ & 0 & 1 & 0 & 0 & 1 & 0 & 0 & 2 & 0 & 3 & 0 & $-2$ & 0 & $-3$& 0  \\\hline
$Z_4$ & 0 & 0 & 1 & 0 & 0 & 1 & 0 & 0 & 2 & 3 & 0 & 0 & $-2$& $-3$ & 0 \\\hline
\end{tabular}

\renewcommand{\arraystretch}{1.0}
\end{table}

Also in the case of \FUTB\ the symmetry is enhanced after the reduction
of couplings.  The superpotential has now a
  $Z_4\times Z_4\times Z_4$ symmetry with charges as shown in Table~\ref{tableB} and  with the
following superpotential
\begin{gather*}
W_B = \sum_{i=1}^{3} \left[ \frac{1}{2}g_{i}^{u}
 {\bf 10}_i{\bf 10}_i H_{i}+
g_{i}^{d} {\bf 10}_i \overline{\bf 5}_{i}
\overline{H}_{i}\right] +
g_{23}^{u} {\bf 10}_2{\bf 10}_3 H_{4} \nonumber\\
\phantom{W_B =}{}   +g_{23}^{d} {\bf 10}_2 \overline{\bf 5}_{3}
\overline{H}_{4}+
g_{32}^{d} {\bf 10}_3 \overline{\bf 5}_{2}
\overline{H}_{4}+
g_{2}^{f} H_{2}
{\bf 24} \overline{H}_{2}+ g_{3}^{f} H_{3}
{\bf 24} \overline{H}_{3}+
\frac{g^{\lambda}}{3} ({\bf 24})^3.
\end{gather*}
For this model the non-degenerate and isolated solutions to
$\gamma^{(1)}_{i}=0$ give us:
\begin{gather}
  (g_{1}^{u})^2
=\frac{8}{5}  g^2, \qquad \big(g_{1}^{d}\big)^2
=\frac{6}{5}g^2, \qquad
(g_{2}^{u})^2=(g_{3}^{u})^2=\frac{4}{5}g^2,\nonumber\\
 \big(g_{2}^{d}\big)^2 = \big(g_{3}^{d}\big)^2=\frac{3}{5}g^2,\qquad
(g_{23}^{u})^2 =\frac{4}{5}g^2, \qquad
\big(g_{23}^{d}\big)^2=\big(g_{32}^{d}\big)^2=\frac{3}{5} g^2,
\nonumber\\
 \big (g^{\lambda}\big)^2 =\frac{15}{7}g^2,\qquad \big(g_{2}^{f}\big)^2
=\big(g_{3}^{f}\big)^2=\frac{1}{2}g^2,\qquad  \big(g_{1}^{f}\big)^2=0,\qquad
\big(g_{4}^{f}\big)^2=0,\label{zoup-SOL52}
\end{gather}
and from the sum rule we obtain \cite{Kobayashi:1997qx}:
\begin{gather*}
m^{2}_{H_u}+
2  m^{2}_{{\bf 10}} =M^2, \qquad
m^{2}_{H_d}-2m^{2}_{{\bf 10}}=-\frac{M^2}{3},\qquad
m^{2}_{\overline{{\bf 5}}}+
3m^{2}_{{\bf 10}}=\frac{4M^2}{3},
\end{gather*}
i.e., in this case we have only two free parameters
$m_{{\bf 10}}\equiv m_{{\bf 10}_3}$  and $M$ for the dimensionful sector.

As already mentioned, after the $SU(5)$ gauge symmetry breaking we
assume we have the MSSM, i.e.\ only two Higgs doublets.  This can be
achieved by introducing appropriate mass terms that allow to perform a
rotation of the Higgs sector \cite{Leon:1985jm, Kapetanakis:1992vx,Mondragon:1993tw,Hamidi:1984gd,
  Jones:1984qd}, in such a~way that only one pair of Higgs doublets,
coupled mostly to the third family, remains light and acquire vacuum
expectation values.  To avoid fast proton decay the usual f\/ine tuning
to achieve doublet-triplet splitting is performed.  Notice that,
although similar, the mechanism is not identical to minimal~$SU(5)$,
since we have an extended Higgs sector.

Thus, after the gauge symmetry of the GUT theory is broken we are left
with the MSSM, with the boundary conditions for the third family given
by the f\/initeness conditions, while the other two families are basically
decoupled.

We will now examine the phenomenology of such all-loop Finite Unif\/ied
Theories with $SU(5)$ gauge group and, for the reasons expressed
above,   we will concentrate only on the
third generation of quarks and leptons. An extension to three
families, and the generation of quark mixing angles and masses in
Finite Unif\/ied Theories has been addressed in~\cite{Babu:2002in},
where several examples are given. These extensions are not considered
here.

\subsection{Restrictions from low-energy observables}
\label{sec:ewpo}

Since the gauge symmetry is spontaneously broken below $M_{\rm GUT}$,
the f\/initeness conditions do not restrict the renormalization
properties at low energies, and all it remains are boundary conditions
on the gauge and Yukawa couplings (\ref{zoup-SOL5}) or
(\ref{zoup-SOL52}), the $h=-MC$ relation (\ref{hY}), and the soft
scalar-mass sum rule (\ref{sumr}) at $M_{\rm GUT}$, as applied in
the two models.  Thus we examine the evolution of these parameters
according to their RGEs up to two-loops for dimensionless parameters
and at one-loop for dimensionful ones with the relevant boundary
conditions.  Below $M_{\rm GUT}$ their evolution is assumed to be
governed by the MSSM.  We further assume a unique supersymmetry
breaking scale $M_{\rm SUSY}$ (which we def\/ine as the geometrical average
of the stop masses) and therefore below that scale the ef\/fective
theory is just the SM.  This allows to evaluate observables at or
below the electroweak scale.

In the following, we brief\/ly describe the low-energy observables used in
our analysis. We discuss the current precision of
the experimental results and the theoretical predictions.
We~also give relevant details of the higher-order perturbative
corrections that we include.
We do not discuss theoretical
uncertainties from the RG running between the high-scale parameters
and the weak scale.
At present, these uncertainties are expected to be
less important than the experimental and theoretical uncertainties of
the precision observables.

As precision observables we f\/irst discuss the 3rd generation quark
masses that are leading to the strongest constraints on the models under
investigation. Next we apply $B$~physics and Higgs-boson mass
constraints.  We also brief\/ly discuss the anomalous magnetic moment of the
muon.

\subsection{Predictions}

\begin{figure}[t]
           \centerline{\includegraphics[width=10cm]{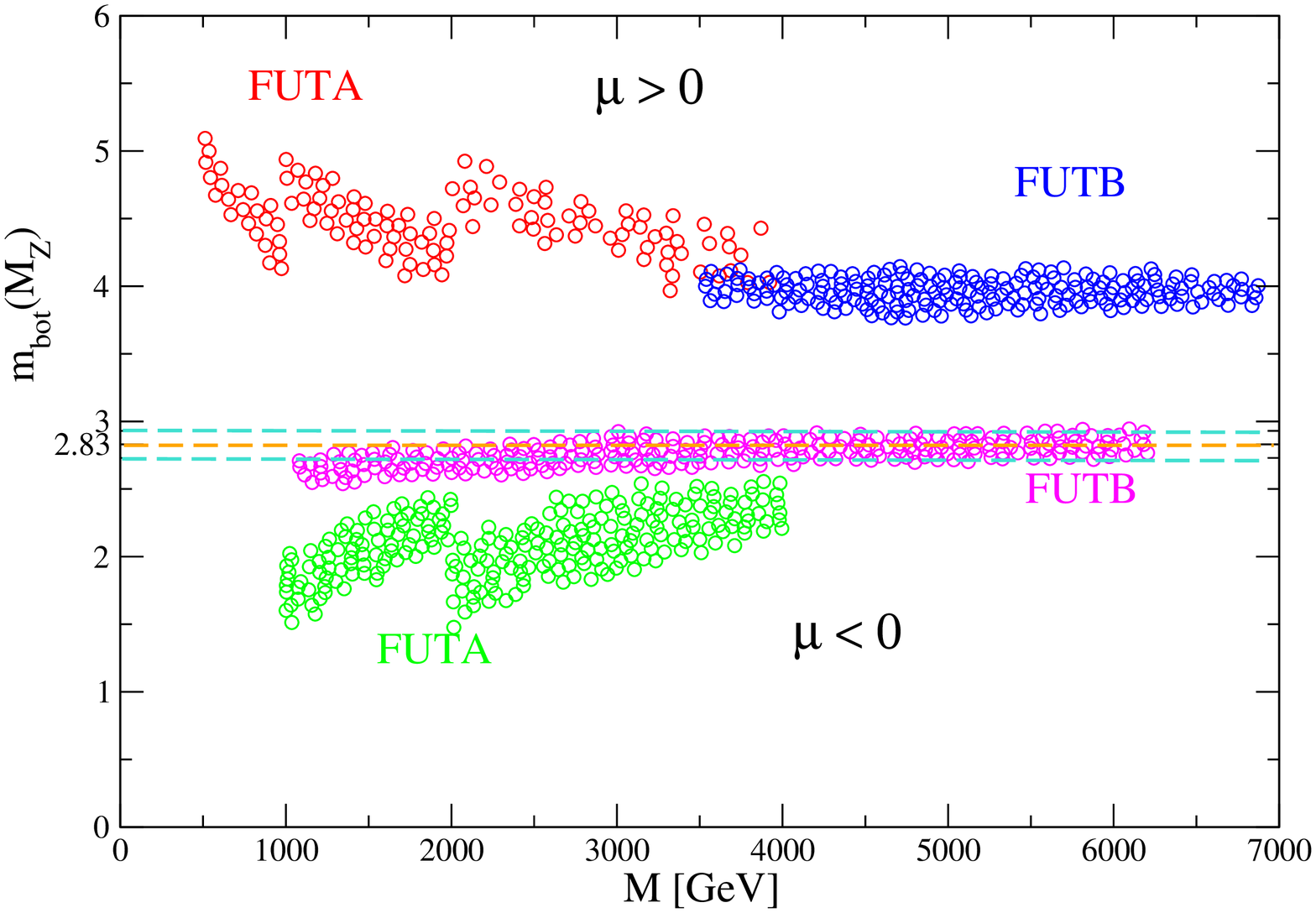}}
\vspace{2mm}

           \centerline{\includegraphics[width=10cm]{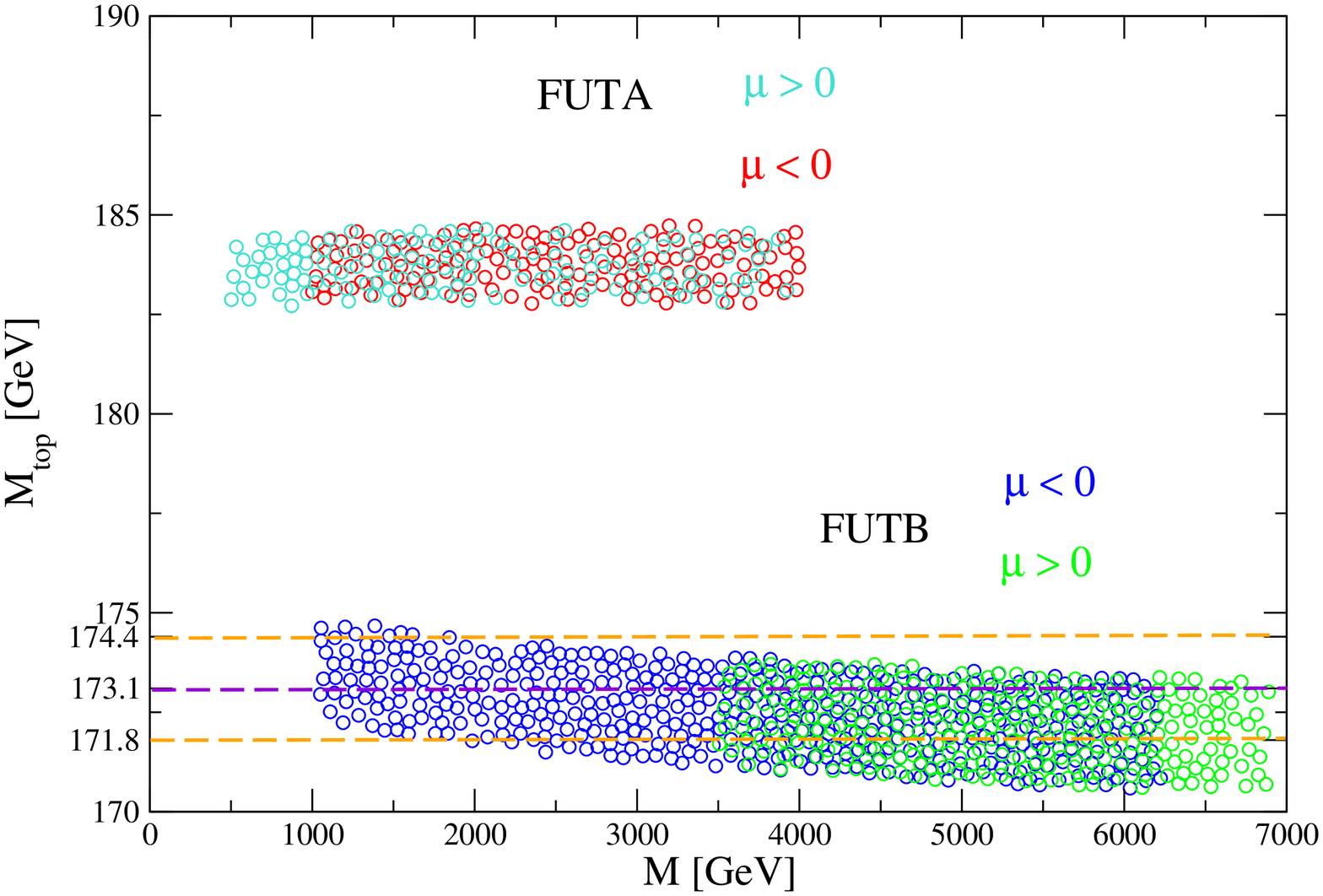}}

       \caption{The bottom quark mass at the $Z$~boson scale (upper)
                and top quark pole mass (lower) are shown
                as function of $M$ for both models.}\label{fig:MtopbotvsM}
\end{figure}

We now present the comparison of the predictions of the four models with
the experimental data, see~\cite{Heinemeyer:2007tz} for more details,
starting with the heavy quark masses.
In Fig.~\ref{fig:MtopbotvsM} we show the {\bf FUTA} and {\bf FUTB}
predictions for the top pole mass, $M_{\rm top}$, and the running bottom
mass at the scale~$M_Z$, $m_{\rm bot}(M_Z)$,
as a function of the  unif\/ied gaugino mass $M$, for the two cases
$\mu <0$ and $\mu >0$.
The running bottom mass is used to avoid the large
QCD uncertainties inherent for the pole mass.
In the evaluation of the bottom mass $m_{\rm bot}$,
we have included the corrections coming from bottom
squark-gluino loops and top squark-chargino loops~\cite{Carena:1999py}.  We
compare the predictions for the running bottom quark mass with the
experimental value,
$m_{b}(M_Z) = 2.83 \pm 0.10~\mathrm{GeV}$~\cite{Amsler:2008zzb}. One can
see that the value of $m_{\rm bot}$ depends
strongly on the sign of $\mu$ due to the above mentioned
radiative corrections involving SUSY particles.
For both models~${\bf A}$ and~${\bf B}$ the values
for $\mu >0$ are above the central experimental value, with
$m_{\rm bot}(M_Z) \sim 4.0$--5.0~GeV.
For $\mu < 0$, on the other hand, model ${\bf B}$ shows
overlap with the experimentally measured values,
$m_{\rm bot}(M_Z) \sim 2.5$--2.8~GeV.
For model ${\bf A}$ we f\/ind $m_{\rm bot}(M_Z) \sim 1.5$--2.6~GeV, and
there is only a small region of allowed parameter space at large $M$
where we f\/ind agreement with the experimental value at the two~$\sigma$ level.
In summary, the experimental determination of $m_{\rm bot}(M_Z)$ clearly
selects the negative sign of $\mu$.

Now we turn to the top quark mass.
The predictions for the top quark mass $M_{\rm top}$ are $\sim 183$ and
$\sim 172$~GeV in the models ${\bf A}$ and ${\bf B}$
respectively, as shown in the lower plot of Fig.~\ref{fig:MtopbotvsM}.
Comparing these predictions with the most recent experimental value
$m_{t}^{\exp} = (173.1 \pm 1.3)$~GeV~\cite{:2009ec}, and recalling
that the theoretical values for $M_{\rm top}$ may suf\/fer from a
correction of $\sim 4 \%$ \cite{Kubo:1997fi,Kobayashi:2001me,Mondragon:2003bp}, we see that clearly model
${\bf B}$ is singled out.
In addition the value of $\tan \beta$ is found to be $\tan \beta \sim 54$ and
$\sim 48$ for models ${\bf A}$ and ${\bf B}$, respectively.
Thus from the comparison of the predictions of the two models with
experimental data only {\bf FUTB} with $\mu < 0$ survives.

We now analyze the impact of further low-energy observables on the model
{\bf FUTB} with $\mu < 0$.
As  additional constraints we consider the following observables:
the rare $b$~decays ${\rm BR}(b \to s \gamma)$ and ${\rm BR}(B_s \to \mu^+ \mu^-)$,
the lightest Higgs boson mass
as well as the density of cold dark matter in the Universe, assuming it
consists mainly of neutralinos. More details and a~complete set of
references can be found in~\cite{Heinemeyer:2007tz}.

For the branching ratio ${\rm BR}(b \to s \gamma)$, we take
the experimental value estimated by the Heavy Flavour Averaging
Group (HFAG) is~\cite{Barate:1998vz,Chen:2001fja,Koppenburg:2004fz}
\[
{\rm BR}(b \to s \gamma ) = \big(3.55 \pm 0.24 {}^{+0.09}_{-0.10} \pm 0.03\big)
                       \cdot 10^{-4},
\]
where the f\/irst error is the combined statistical and uncorrelated systematic
uncertainty, the latter two errors are correlated systematic theoretical
uncertainties and corrections respectively.

For the branching ratio ${\rm BR}(B_s \to \mu^+ \mu^-)$, the SM prediction is
at the level of $10^{-9}$, while the present
experimental upper limit from the Tevatron is
$4.7 \cdot 10^{-8}$ at the $95\%$ C.L.~\cite{hfag}, still providing the
possibility for the MSSM to dominate the SM contribution.

Concerning the lightest Higgs boson mass, $M_h$, the SM bound of
$114.4$~GeV \cite{Barate:2003sz,Schael:2006cr} can be applied, since the
main SM search channels are not suppressed in {\rm FUTB}. For the
prediction we use the code
{\tt FeynHiggs}
\cite{Heinemeyer:1998yj,Heinemeyer:1998np,Degrassi:2002fi,Frank:2006yh}.

\begin{figure}[t]
           \centerline{\includegraphics[width=10cm]{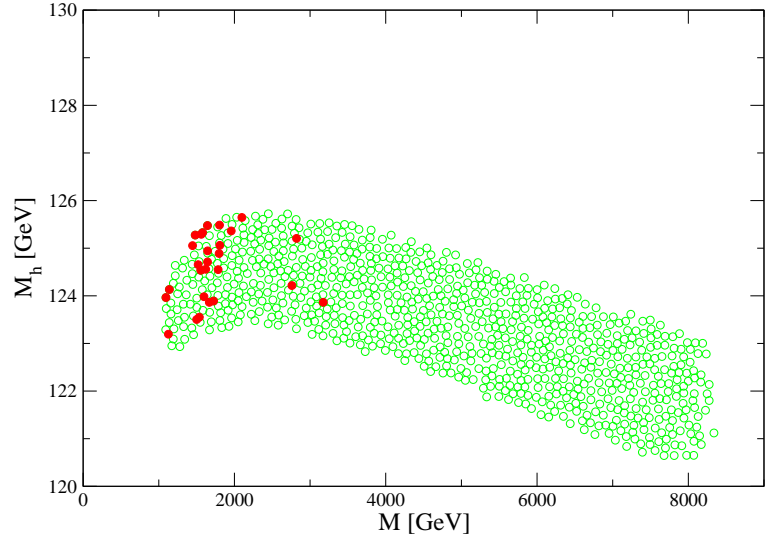}}
        \caption{The lightest Higgs mass, $M_h$,  as function of $M$ for
          the model {\bf FUTB} with $\mu < 0$, see text.}\label{fig:Higgs}
\end{figure}

The prediction of the lightest Higgs boson mass as a function of $M$ is
shown in Fig.~\ref{fig:Higgs}. The light (green) points shown are in agreement
with the two $B$-physics observables listed above.
The lightest Higgs mass ranges in
\begin{gather}
M_h \sim 121{-}126~{\rm GeV} ,
\label{eq:Mhpred}
\end{gather}
where the uncertainty comes from
variations of the soft scalar masses, and
from f\/inite (i.e.~not logarithmically divergent) corrections in
changing renormalization scheme.  To this value one has to add $\pm 3$
GeV coming from unknown higher order corrections~\cite{Degrassi:2002fi}.
We have also included a~small variation,
due to threshold corrections at the GUT scale, of up to 5\% of the
FUT boundary conditions.
Thus, taking into account the $B$~physics constraints
 results naturally in a~light Higgs boson that fulf\/ills
the LEP bounds~\cite{Barate:2003sz,Schael:2006cr}.

In the same way the whole SUSY particle spectrum can be derived.
The resulting SUSY masses for {\bf FUTB} with $\mu < 0$ are rather large.
The lightest SUSY particle starts around 500~GeV, with the rest
of the spectrum being very heavy. The observation of SUSY particles at
the LHC or the ILC will only be possible in very favorable parts of the
parameter space. For most parameter combination only a SM-like light
Higgs boson in the range of equation~(\ref{eq:Mhpred}) can be observed.

We note that with such a heavy SUSY spectrum
the anomalous magnetic moment of the muon, \mbox{$(g-2)_\mu$}
(with $a_\mu \equiv (g-2)_\mu/2$), gives only a negligible correction to
the SM prediction.
The comparison of the experimental result and the SM value (based on the
latest combination using $e^+e^-$ data)~\cite{Davier:2009zi}
\[
a_\mu^{\exp}-a_\mu^{\rm theo} = (24.6 \pm 8.0) \cdot 10^{-10}
\]
would disfavor {\bf FUTB} with $\mu < 0$ by about $3 \sigma$. However,
since the SM is not regarded as excluded by $(g-2)_\mu$,
we still see {\bf FUTB} with $\mu < 0$ as the only surviving model.

Further restrictions on the parameter space can arise from the
requirement that the lightest SUSY particle (LSP) should give the
right amount of cold dark matter (CDM) abundance. The LSP should be
color neutral, and the lightest neutralino appears to be a suitable
candidate~\cite{Goldberg:1983nd,Ellis:1983ew}.
In the case where
all the soft scalar masses are universal at the unf\/ication scale,
there is no region of $M$ below ${\cal O}$(few~TeV) in which
$m_{\tilde \tau} > m_{\chi^0}$ is satisf\/ied,
where $m_{\tilde \tau}$ is the lightest~$\tilde \tau$ mass,
and $m_{\chi^0}$ the lightest neutralino mass. An electrically charged
LSP, however, is not in agreement with CDM searches. But once the
universality condition is relaxed this problem can be solved
naturally, thanks to the sum rule (\ref{sumr}).  Using this equation
a comfortable parameter space is found for
{\bf FUTB} with $\mu < 0$ (and also for {\bf FUTA} and both signs of $\mu$).
that fulf\/ills the conditions of (a)~successful radiative
electroweak symmetry breaking, (b)~$m_{\tilde\tau}> m_{\chi^0}$.

Calculating the CDM abundance in these FUT models one f\/inds that usually
it is very large, thus a
mechanism is needed in our model to reduce it. This issue could, for
instance, be related to another problem, that of neutrino masses.
This type of masses cannot be generated naturally within the class of
f\/inite unif\/ied theories that we are considering in this paper,
although a non-zero value for neutrino masses has clearly been
established~\cite{Amsler:2008zzb}.  However, the class of FUTs
discussed here can, in principle, be easily extended by introducing
bilinear $R$-parity violating terms that preserve f\/initeness and
introduce the desired neutrino masses \cite{Valle:1998bs}.  $R$-parity
violation~\cite{Dreiner:1997uz,Bhattacharyya:1997vv,Allanach:1999ic,Romao:1991ex}
would have a small impact on the collider phenomenology discussed here
(apart from fact the SUSY search strategies could not rely on a
``missing energy'' signature), but remove the CDM bound completely.  The
details of such a possibility in the present framework attempting to
provide the models with realistic neutrino masses will be discussed
elsewhere.  Other mechanisms, not involving $R$-parity violation (and
keeping the ``missing energy'' signature), that could be invoked if the
amount of CDM appears to be too large, concern the cosmology of the
early universe.  For instance, ``thermal
inf\/lation''~\cite{Lyth:1995ka} or ``late time entropy
injection''~\cite{Gelmini:2006pw} could bring the CDM density into
agreement with the WMAP measurements.  This kind of modif\/ications of
the physics scenario neither concerns the theory basis nor the
collider phenomenology, but could have a strong impact on the CDM
derived bounds.

\begin{figure}[t]
       \centerline{\includegraphics[width=10cm,angle=0]{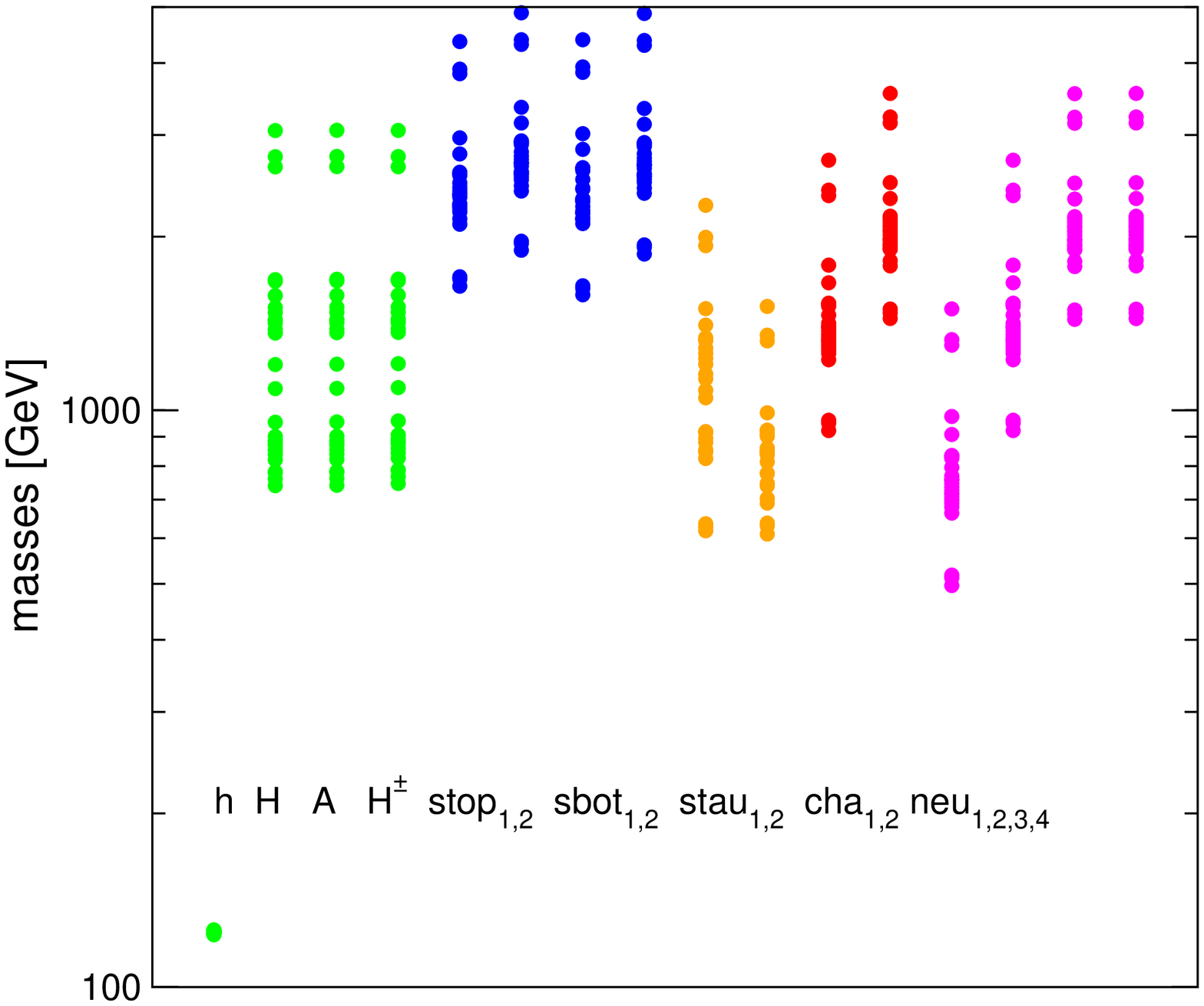}}
           \caption{The particle spectrum of model {\bf FUTB} with $\mu <0$,
             where
             the points shown are in agreement with the quark mass
             constraints, the
             $B$-physics observables and the loose CDM constraint.
  The light (green) points on the left are the various Higgs boson masses. The
  dark (blue) points following are the two scalar top and bottom masses,
  followed by lighter (beige) scalar tau masses. The darker (red) points to
  the right are the two chargino masses followed by the lighter shaded (pink)
  points indicating the neutralino masses.}
\label{fig:masses}
\vspace{2mm}
 \centerline{\includegraphics[width=10cm,angle=0]{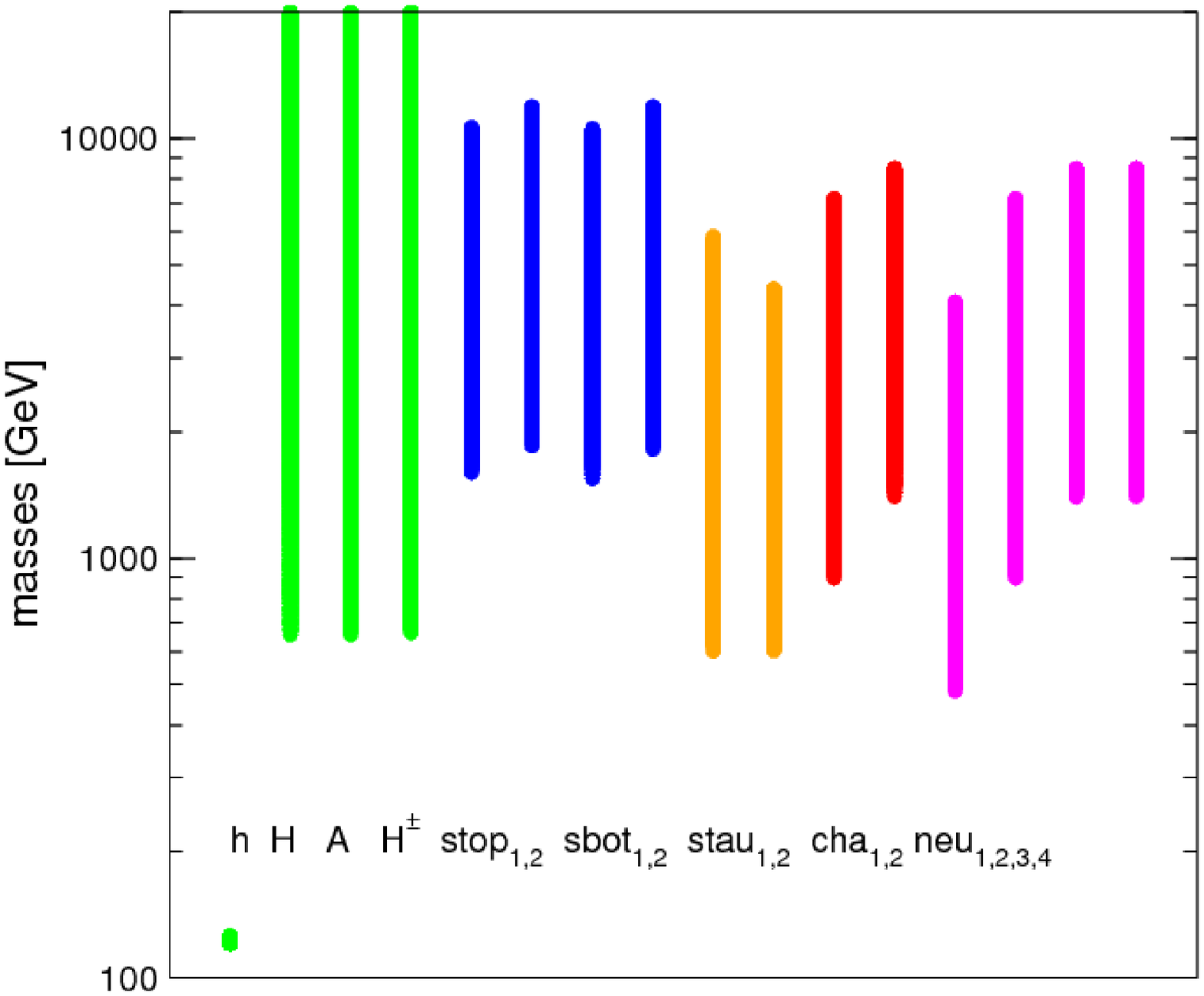}}
           \caption{The particle spectrum of model {\bf FUTB} with $\mu <0$,
             the points shown are in agreement with the quark mass
             constraints, the
             $B$-physics observables, but the loose CDM constraint has been
             omitted (motivated by possible $R$-parity violation, see text).
  The color coding is as in Fig.~\ref{fig:masses}.}\label{fig:masses2}
\end{figure}

Therefore, in order to get an impression of the
{\em possible} impact of the CDM abundance on the collider phenomenology
in our models under investigation, we will analyze the case that the LSP
does contribute to the CDM density, and apply a more loose bound of
\begin{gather}
\Omega_{\rm CDM} h^2 < 0.3.
\label{cdmloose}
\end{gather}
(Lower values than the ones permitted by \eqref{cdmloose} are
naturally allowed if another particle than the lightest neutralino
constitutes CDM.)  For our evaluation we have used the code {\tt
  MicroMegas}~\cite{Belanger:2001fz,Belanger:2004yn}.
The prediction for the lightest Higgs mass, $M_h$ as
function of $M$ for the model {\bf FUTB} with $\mu < 0$
is shown in Fig.~\ref{fig:Higgs}. The dark (red) dots are the points that pass
the constraints in \eqref{cdmloose} (and that have the lightest neutralino as
LSP), which favors relatively light values of~$M$.
The full particle
spectrum of model {\bf FUTB} with $\mu <0$, again compliant with quark mass
constraints, the $B$ physics observables (and with the loose CDM constraint),
is shown in Fig.~\ref{fig:masses}. The masses of the particles increase with
increasing values of the unif\/ied gaugino mass $M$.
One can see that large parts of the spectrum are in the kinematic reach of the
LHC. A numerical example of such a light spectrum is shown in
Table~\ref{table:mass}. The colored part of this spectrum as well as the
lightest Higgs boson should be (relatively easily) accessible at the LHC.

\begin{table}[t]
\centering

\caption{A representative spectrum of a light {\bf FUTB}, $\mu <0$ spectrum.}
\label{table:mass}

\vspace{1mm}

\begin{tabular}{|l|l||l|l|}
\hline
Mbot($M_Z$) &  2.71 GeV &
Mtop &    172.2 GeV\\ \hline
Mh &  123.1 GeV & 
MA &  680 GeV\\ \hline
MH &  679 GeV&
MH$^\pm$ &  685 GeV \\ \hline 
 Stop1 &  1876 GeV &
Stop2 &    2146 GeV \\ \hline
Sbot1 &   1849 GeV & 
Sbot2 &    2117 GeV\\ \hline
Mstau1 &    635 GeV &
Mstau2 &    867 GeV\\ \hline
Char1 &    1072 GeV &
Char2 &    1597 GeV\\ \hline
Neu1  &    579 GeV &
Neu2  &    1072 GeV \\ \hline
Neu3  &    1591 GeV &
Neu4  &    1596 GeV \\ \hline
 M1 &    580 GeV&
 M2 &   1077 GeV\\ \hline
 Mgluino &    2754 GeV &
&  \\
\hline
\end{tabular}

\end{table}

Finally for the model {\bf FUTB} with $\mu < 0$ we show the particle spectrum,
where only the quark mass constraints and the $B$ physics observables are
taken into account. The loose CDM constraint, on the other hand, has been
omitted, motivated by the possible $R$-parity violation as discussed above.
Consequently, following Fig.~\ref{fig:Higgs}, larger values of $M$ are allowed,
resulting in a~heavier spectrum, as can be seen in Fig.~\ref{fig:masses2}. In
this case only part of the parameter space would be in the kinematic reach of
the LHC. Only the SM-like light Higgs boson remains observable for the whole
parameter range.

A more detailed analysis can be found in \cite{Heinemeyer:2007tz}.

\section[Finite ${SU(3)^3}$ model]{Finite $\boldsymbol{{SU(3)^3}}$ model}

We now examine the possibility of constructing realistic FUTs based on
product gauge groups. Consider an $N=1$ supersymmetric theory, with
gauge group $SU(N)_1 \times SU(N)_2 \times \cdots \times SU(N)_k$,
with $n_f$ copies (number of families) 
of the
supersymmetric multiplets $(N,N^*,1,\dots,1) + (1,N,N^*,\dots,1) +
\cdots + (N^*,1,1,\dots,N)$.  The one-loop $\beta$-function
coef\/f\/icient in the renor\-ma\-li\-za\-tion-group equation of each $SU(N)$
gauge coupling is simply given by
\begin{gather}
b = \left( -\frac{11}{3} + \frac{2}{3} \right) N + n_f \left( \frac{2}{3}
 + \frac{1}{3} \right) \left( \frac{1}{2} \right) 2 N = -3 N + n_f
N.
\label{3gen}
\end{gather}
This means that $n_f = 3$ is the only solution of equation \eqref{3gen} that
yields $b = 0$.  Since $b=0$ is a~necessary condition for a f\/inite f\/ield theory, the existence of three
families of quarks and leptons is natural in such models, provided the
matter content is exactly as given above.

The model of this type with best phenomenology is the $SU(3)^3$ model
discussed in~\cite{Ma:2004mi}, where the details of the model are
given. It corresponds to the well-known example of $SU(3)_C
\times SU(3)_L \times
SU(3)_R$~\cite{Derujula:1984gu,Lazarides:1993sn,Lazarides:1993uw,Ma:1986we},
 with quarks transforming as
\begin{gather}
  q = \begin{pmatrix} d & u & h \\ d & u & h \\ d & u & h \end{pmatrix}
\sim (3,3^*,1), \qquad
    q^c = \begin{pmatrix} d^c & d^c & d^c \\ u^c & u^c & u^c \\ h^c & h^c & h^c
\end{pmatrix}
    \sim (3^*,1,3),
\label{2quarks}
\end{gather}
and leptons transforming as
\begin{gather}
\lambda = \begin{pmatrix} N & E^c & \nu \\ E & N^c & e \\ \nu^c & e^c & S
\end{pmatrix}
\sim (1,3,3^*).
\label{3leptons}
\end{gather}
Switching the f\/irst and third rows of $q^c$ together with the f\/irst and
third columns of $\lambda$, we obtain the alternative left-right model f\/irst
proposed in~\cite{Ma:1986we} in the context of superstring-inspired~$E_6$.

In order for all the gauge couplings to be equal at $M_{\rm GUT}$, as is
suggested by the LEP results~\cite{Amaldi:1991cn}, the cyclic symmetry $Z_3$
must be imposed, i.e.
\begin{gather*}
q \to \lambda \to q^c \to q,
\end{gather*}
where $q$ and $q^c$ are given in equation~(\ref{2quarks}) and $\lambda$ in
equation~(\ref{3leptons}).  Then,
the f\/irst of the f\/initeness conditions (\ref{1st}) for one-loop
f\/initeness, namely the vanishing of the gauge $\beta$-function is
satisf\/ied.

Next let us consider the second condition, i.e.\ the vanishing of the
anomalous dimensions of all superf\/ields, equation~(\ref{2nd}).  To do that
f\/irst we have to write down the superpotential.  If there is just one
family, then there are only two trilinear invariants, which can be
constructed respecting the symmetries of the theory, and therefore can
be used in the superpotential as follows
\begin{gather*}
f \,{\rm Tr} (\lambda q^c q) + \frac{1}{6} f'  \epsilon_{ijk} \epsilon_{abc}
(\lambda_{ia} \lambda_{jb} \lambda_{kc} + q^c_{ia} q^c_{jb} q^c_{kc} +
q_{ia} q_{jb} q_{kc}),
\end{gather*}
where $f$ and $f'$ are the Yukawa couplings associated to each invariant.
%
Quark and leptons obtain masses when the scalar parts of the
superf\/ields $(\tilde N,\tilde N^c)$ obtain vacuum expectation values (vevs),
\begin{gather*}
m_d = f \langle \tilde N \rangle, \qquad m_u = f \langle \tilde N^c \rangle, \qquad
m_e = f' \langle \tilde N \rangle, \qquad m_\nu = f' \langle \tilde N^c \rangle.
\end{gather*}

With three families, the most general superpotential contains 11 $f$
couplings, and 10 $f'$ couplings, subject to 9 conditions, due to the
vanishing of the anomalous dimensions of each superf\/ield.  The
conditions are the following
\begin{gather}
\sum_{j,k} f_{ijk} (f_{ljk})^* + \frac{2}{3} \sum_{j,k} f'_{ijk}
(f'_{ljk})^* = \frac{16}{9} g^2 \delta_{il} ,
\label{19}
\end{gather}
where
\begin{gather*}
  f_{ijk} = f_{jki} = f_{kij}, \qquad 
  f'_{ijk} = f'_{jki} = f'_{kij} = f'_{ikj} = f'_{kji} = f'_{jik}.
\end{gather*}
Quarks and leptons receive  masses when  the scalar part of the
superf\/ields $\tilde N_{1,2,3}$ and $\tilde N^c_{1,2,3}$ obtain vevs as follows
\begin{gather*}
  ({\cal M}_d)_{ij} = \sum_k f_{kij} \langle \tilde N_k \rangle, \qquad
   ({\cal M}_u)_{ij} = \sum_k f_{kij} \langle \tilde N^c_k \rangle, \\ 
  ({\cal M}_e)_{ij} = \sum_k f'_{kij} \langle \tilde N_k \rangle, \qquad
   ({\cal M}_\nu)_{ij} = \sum_k f'_{kij} \langle \tilde N^c_k \rangle.
\end{gather*}

We will assume that the below $M_{\rm GUT}$ we have the usual MSSM, with
the two Higgs doublets coupled maximally to the third generation.
Therefore we have to choose
the linear combinations $\tilde N^c = \sum_i a_i \tilde N^c_i$ and
$\tilde N = \sum_i b_i \tilde N_i$ to play the role of the two Higgs
doublets, which will be responsible for the electroweak symmetry
breaking.  This can be done by choosing appropriately the masses in
the superpotential \cite{Leon:1985jm}, since they are not
constrained by the f\/initeness conditions.  We choose that
the two Higgs doublets are predominately coupled to the third
generation. Then these two Higgs doublets couple to the three
families dif\/ferently, thus providing the freedom to understand
their dif\/ferent masses and mixings.
The remnants of the $SU(3)^3$ FUT are the boundary conditions on the
gauge and Yukawa couplings, i.e.~\eqref{19}, the $h=-MC$
relation, and the soft scalar-mass sum rule equation~(\ref{sumr}) at $M_{\rm
  GUT}$, which, when applied to the present model, takes the form
\[
m^2_{H_u} + m^2_{\tilde t^c} + m^2_{\tilde q} = M^2 =
m^2_{H_d} + m^2_{\tilde b^c} + m^2_{\tilde q},
\]
where   ${\tilde t^c}$, ${\tilde b^c}$, and ${\tilde q}$ are the
scalar parts of the corresponding

Concerning the solution to equation \eqref{19} we consider two versions of the model:

I) An all-loop f\/inite model with a unique and isolated solution, in
which $f'$ vanishes, which leads to the following relation
\begin{gather*}
f^2 = f^2_{111} = f^2_{222} = f^2_{333} = \frac{16}{9} g^2.
\end{gather*}
  As for
the lepton masses, because all $f'$ couplings have been f\/ixed to be
zero at this order, in principle they would be expected to appear
radiatively induced by the scalar lepton masses appearing in the SSB
sector of the theory.  However, due to the f\/initeness
conditions they cannot appear radiatively and remain as a
problem for further study.

II) A two-loop f\/inite solution, in which we keep $f'$ non-vanishing
and we use it to introduce the lepton masses. The model in turn
becomes f\/inite only up to two-loops since the corresponding solution
of equation \eqref{19} is not an isolated one any more, i.e.\ it is a parametric
one.  In this case we have the following boundary conditions for the
Yukawa couplings
\begin{gather*}
f^2 = r \left(\frac{16}{9}\right) g^2,\qquad
f'^2 = (1-r) \left(\frac{8}{3}\right) g^2,
\end{gather*}
where $r$ is a free parameter which parametrizes the dif\/ferent
solutions to the f\/initeness conditions.
As for the boundary conditions of the soft scalars, we have the
universal case.

\subsection[Predictions for  ${SU(3)^3}$]{Predictions for  $\boldsymbol{SU(3)^3}$}\label{section6}

Below $M_{\rm GUT}$ all couplings and masses of the theory run according
to the RGEs of the MSSM.  Thus we examine the evolution of these
parameters according to their RGEs up to two-loops for dimensionless
parameters and at one-loop for dimensionful ones imposing the
corresponding boundary conditions.  We further assume a unique
supersymmetry breaking scale $M_{\rm SUSY}$ and below that scale the
ef\/fective theory is just the SM.

We compare our predictions with the most recent experimental value $
m_t^{\exp} = (173.1 \pm 1.3)$~GeV~\cite{:2009ec}, and recall
that the theoretical values for $m_t$ suf\/fer from a correction of
\mbox{$\sim 4$\%}~\cite{Kubo:1997fi,Kobayashi:2001me,Mondragon:2003bp}. In
the case of the bottom quark, we take again the value evaluated at
$M_Z$, $m_b (M_Z)=2.83\pm 0.10$~GeV~\cite{Amsler:2008zzb}.  In
the case of model I, the predictions for the top quark mass (in this
case $m_b$ is an input) $m_t$ are $\sim 183$~GeV for $\mu < 0 $,
which is above the experimental value, and there are no solutions for $\mu>0$.

For the two-loop model II, we look for the values of the parameter
$r$ which comply with the experimental limits given above for top and
bottom quarks masses. In the case of $\mu >0$, for the bottom quark,
the values of $r$ lie in the range $0.15 \lesssim r \lesssim 0.32$.
For the top mass, the range of values for r is $0.35 \lesssim r
\lesssim 0.6$. From these values we can see that there is a very small
region where both top and bottom quark masses are in the experimental
range for the same value of $r$.  In the case of $\mu<0$ the situation
is similar, although slightly better, with the range of values $0.62
\lesssim r \lesssim 0.77$ for the bottom mass, and $0.4 \lesssim r
\lesssim 0.62$ for the top quark mass.  So far in the analysis, the
masses of the new particles $h$'s and $E$'s of all families were taken
to be at the $M_{\rm GUT}$ scale. Taking into account new thresholds for
these exotic particles below $M_{\rm GUT}$ we hope to f\/ind a wider
phenomenologically viable parameter space. The details of the
predictions of the $SU(3)^3$ are currently under a careful re-analysis in
view of the new value of the top-quark mass, the possible new
thresholds for the exotic particles, as well as dif\/ferent intermediate
gauge symmetry breaking into $SU(3)_c \times SU(2)_L \times SU(2)_R \times
U(1)$ \cite{new-HMMZ}.

\section{Conclusions}

A number of proposals and ideas have matured with time and have
survived after careful theoretical studies and confrontation with
experimental data. These include part of the original GUTs ideas,
mainly the unif\/ication of gauge couplings and, separately, the
unif\/ication of the Yukawa couplings, a version of f\/ixed point
behaviour of couplings, and certainly the necessity of supersymmetry
as a way to take care of the technical part of the hierarchy problem.
On the other hand, a very serious theoretical problem, namely the
presence of divergencies in Quantum Field Theories (QFT), although
challenged by the founders of QFT
\cite{Dirac:book,Dyson:1952tj,Weinberg:2009ca}, was mostly forgotten
in the course of developments of the f\/ield partly due to the
spectacular successes of renormalizable f\/ield theories, in particular
of the SM. However, as it was already mentioned in the Introduction,
fundamental developments in Theoretical Particle Physics are based in
reconsiderations of the problem of divergencies and serious attempts
to solve it. These include the motivation and construction of string
and non-commutative theories, as well as $N=4$ supersymmetric f\/ield
theories \cite{Mandelstam:1982cb,Brink:1982wv}, $N=8$ supergravity
\cite{Bern:2009kd,Kallosh:2009jb,Bern:2007hh,Bern:2006kd,Green:2006yu}
and the AdS/CFT correspondence \cite{Maldacena:1997re}.  It is a
thoroughly fascinating fact that many interesting ideas that have
survived various theoretical and phenomenological tests, as well as
the solution to the UV divergencies problem, f\/ind a common ground in
the framework of $N=1$ Finite Unif\/ied Theories, which we have
described in the previous sections. From the theoretical side they
solve the problem of UV divergencies in a minimal way. On the
phenomenological side, since they are based on the principle of
reduction of couplings (expressed via RGI relations among couplings
and masses), they provide strict selection rules in choosing realistic
models which lead to testable predictions. The celebrated success of
predicting the top-quark mass
\cite{Kapetanakis:1992vx,Mondragon:1993tw,Kubo:1994bj,Kubo:1994xa,Kubo:1995zg,Kubo:1996js}
is now extented to the predictions of the Higgs masses and the
supersymmetric spectrum of the MSSM. At least the prediction of the
lightest Higgs sector is expected to be tested in the next couple of
years at LHC.

\subsection*{Acknowledgements}
It is a pleasure for one of us (G.Z.) to thank the Organizing
Committee for the very warm hospitality of\/fered.  This work is
partially supported by the NTUA's basic research support programme
2008 and 2009, and the European Union's RTN programme under contract
MRTN-CT-2006-035505.  Supported also by a mexican PAPIIT grant
IN112709, and by Conacyt grants 82291 and 51554-F.

\pdfbookmark[1]{References}{ref}
\LastPageEnding

\end{document}